\documentclass[aps,prb,floatfix,twocolumn,superscriptaddress]{revtex4}

\usepackage{mathrsfs}
\usepackage{graphicx}
\usepackage[english]{babel}
\usepackage{amsmath}
\usepackage{amssymb}
\usepackage{xcolor}
\usepackage{cancel}
\usepackage[normalem]{ulem}

\usepackage{relsize}
\usepackage{CJK}
\usepackage{dsfont}
\usepackage{bm}% bold math

\newcommand{\me}{\mathrm{e}}
\newcommand{\mi}{\mathrm{i}}

\newcommand{\dif}{\mathrm{d}}

\allowdisplaybreaks

\begin{document}

\title{Thermal Suppression of Dynamical Quantum Phase Transitions in Finite-Dimensional Systems\\
{ A Quasi-Hermitian Framework}}

\author{Jia-Chen Tang}
\affiliation{School of Physics, Southeast University, Jiulonghu Campus, Nanjing 211189, China}

\author{Xu-Yang Hou}
\affiliation{School of Physics, Southeast University, Jiulonghu Campus, Nanjing 211189, China}

\author{Ming-Zhang Wang}
\affiliation{School of Physics, Southeast University, Jiulonghu Campus, Nanjing 211189, China}

\author{Hao Guo}
\email{guohao.ph@seu.edu.cn}
\affiliation{School of Physics, Southeast University, Jiulonghu Campus, Nanjing 211189, China}
\affiliation{Hefei National Laboratory, Hefei 230088, China}

\begin{abstract}
We investigate dynamical quantum phase transitions (DQPTs) in finite-dimensional systems prepared in thermal equilibrium states and subjected to a sudden quench.
A mixed-state Loschmidt amplitude is constructed from first principles within a metric-stationary pseudo-Hermitian framework, providing a self-contained derivation of the finite-temperature quench dynamics.
Applying this framework to an $N$-level model consisting of a two-level sector coupled to $N-2$ spectator states, we find that temperature controls the DQPTs through the redistribution of thermal weights among the eigenstates.
This mechanism leads to a dimensionality-dependent threshold temperature that becomes finite when the Hilbert-space dimension reaches five, above which the Loschmidt amplitude loses all real zeros and the DQPTs are fully suppressed.
The thermal suppression mechanism suggests a general principle for controlling dynamical criticality through thermal occupation, while the quasi-Hermitian framework provides the self-consistent foundation for its rigorous derivation.
\end{abstract}
\maketitle

% ============================================================
\section{Introduction}

Dynamical quantum phase transitions (DQPTs) are a fundamental concept in nonequilibrium quantum many-body physics.
They manifest as nonanalytic behavior in the real-time evolution of quantum systems following a sudden quench of a Hamiltonian parameter~\cite{Heyl2013, Heyl2018}.
The central diagnostic tool is the Loschmidt amplitude $\mathcal{G}(t) = \langle \psi(0)|\psi(t)\rangle$, whose zeros at critical times $t_n^*$ define the DQPTs.
The corresponding rate function $r(t) = -\ln |\mathcal{G}(t)|^2$ develops cusp-like singularities at these critical times, in close analogy with equilibrium phase transitions~\cite{Heyl2014, Heyl2015, Budich2016, Vajna2015a, Vajna2015b}.

This framework has been extended to mixed states via purification of the density matrix, yielding a generalized Loschmidt amplitude $\mathcal{G}(t) = \mathrm{Tr}[\rho(0)\me^{-\mi Ht}]$ that naturally incorporates finite temperature~\cite{Bhattacharya2017, Heyl2017, Hou2020a, Hou2022, Sedlmayr2018, Mera2018, Lang2018a, Lang2018b}.
Subsequent studies have broadened DQPTs to nonintegrable models, long-range interactions, Floquet-driven systems, and other settings~\cite{Halimeh2017, Homrighausen2017, Zunkovic2018, Karrasch2013, Karrasch2014, Andraschko2014, Schmitt2015, Sharma2015, Divakaran2016, Zamani2020, Jafari2021a, Jafari2021b, Jafari2022a, Naji2022, Yang2019, Zhou2021a, Hamazaki2021}.

In parallel, non-Hermitian quantum systems have attracted intense interest owing to their rich phenomenology in open systems and systems with gain and loss~\cite{Ashida2020, El-Ganainy2018, Bender1998, Bergholtz2021, Gong2018, Kunst2018, Yao2018, Kawabata2019}.
Non-Hermitian Hamiltonians generally possess non-orthogonal eigenstates, requiring a biorthogonal basis for spectral decomposition~\cite{Brody2014}.
This structural difference profoundly affects DQPTs, and a complete framework for biorthogonal DQPTs has recently been developed for pure states~\cite{Tang2022Biorthogonal, Sun2022Biorthogonal}, revealing exotic features such as half-integer jumps in the dynamical topological order parameter~\cite{Mondal2022, Mondal2023PhysRevB, Zhou2018, Zhou2021b}.
Extending these concepts to mixed states requires a consistent treatment of the inner product, normalization, and probabilistic interpretation of the Loschmidt echo, which remains an open challenge.

A natural resolution is provided by quasi-Hermitian quantum mechanics, where a positive-definite metric operator defines a physical Hilbert space in which the Hamiltonian is self-adjoint and the spectrum is real~\cite{Mostafazadeh2002, Mostafazadeh2010, Bender2007, Das2011, Zhang2020}.
In a generic quasi-Hermitian quench, the initial and final Hamiltonians may correspond to different metrics, complicating the definition of the Loschmidt amplitude.
The fixed-metric scenario, where both Hamiltonians share the same metric, avoids these complications and preserves norm conservation throughout the quench~\cite{Guo2020, Hou2021, Galindo2021, Zhang2021}, providing a clean setting in which to isolate the effects of temperature and Hilbert-space dimensionality.

The interplay of temperature and non-Hermiticity in DQPTs has been explored in several two-band models~\cite{Mondal2023PhysRevB}, where thermal fluctuations were shown to wash out the half-integer winding numbers observed at zero temperature.
However, a systematic, first-principles derivation of the mixed-state Loschmidt amplitude for quasi-Hermitian systems is still lacking.
Existing works often postulate the form of $\mathcal{G}(t)$ by analogy with the Hermitian case, rather than constructing it from the underlying metric structure, biorthogonal eigenstates, and purification formalism.
In this work, we fill this gap by providing a complete and self-contained derivation of the finite-temperature quench dynamics within a fixed-metric framework.
We focus on the metric-stationary scenario as the necessary first step: only after the fixed-metric framework is firmly established can one meaningfully address the more exotic metric-changing quenches, where genuinely non-Hermitian signatures such as complex interference terms and parameter-space holonomies are expected to emerge.

Our primary objective is to investigate how temperature governs DQPTs through the redistribution of thermal weights among the eigenstates.
We show that this mechanism becomes increasingly effective as the Hilbert-space dimension grows, and for systems with five or more levels it leads to a finite threshold temperature above which DQPTs are completely suppressed.
A detailed analysis further reveals that, under the metric-stationary condition, the non-Hermitian parameters are absorbed into a global similarity transformation and manifest only as a renormalization of energy scales.
The thermal suppression mechanism is therefore a universal feature of finite-dimensional multi-level systems, while the quasi-Hermitian framework provides the self-consistent foundation for its rigorous derivation.

The paper is organized as follows.
Section~\ref{Sec2} develops the metric-stationary quench framework: it reviews the pseudo-Hermitian formalism, extends the dynamics to mixed states, introduces the purification structure, and derives the mixed-state Loschmidt amplitude from first principles.
Section~\ref{Sec3} applies this framework to an $N$-level $\mathcal{PT}$-symmetric model with spectator states, where the thermal suppression mechanism is analyzed for two quench protocols and the emergence of a finite threshold temperature for $N\ge 5$ is demonstrated.
Section~\ref{Sec4} examines the role of non-Hermiticity and establishes the strict equivalence between the quasi-Hermitian Loschmidt amplitude and that of the associated Hermitian system under a fixed metric.
We conclude in Sec.~\ref{Sec5} with a summary and outlook.

% ============================================================
\section{Pseudo-Hermitian Framework and Quench Dynamics}\label{Sec2}

\subsection{Spectral properties of pseudo-Hermitian operators}\label{Sec2A}

We begin by recalling the essential properties of quasi-Hermitian systems that are necessary for the subsequent formulation of mixed-state dynamics.

A Hamiltonian $H$ is called pseudo-Hermitian if there exists a linear, Hermitian, and invertible operator $\eta$ such that $H^\dagger = \eta H \eta^{-1}$.
When $\eta$ is chosen to be positive definite, it defines a genuine physical inner product $(\phi,\psi)_{\eta} := \langle \phi | \eta | \psi \rangle$, which turns the state space into a true Hilbert space $(\mathcal H,(\cdot,\cdot)_{\eta})$.  In this case the eigenvalues of $H$ are guaranteed to be real, and $H$ is called quasi-Hermitian.  With respect to the physical inner product, we introduce the $\eta$-adjoint operation $A^{\ddagger} := \eta^{-1} A^\dagger \eta$, which has been successfully employed in the our earlier analysis of mixed-state geometric phases of quasi-Hermitian quantum systems~\cite{Hou2026UhlmannQuasiHermitian}.  An operator $O$ satisfying $O^{\ddagger}=O$ is self-adjoint under the $\eta$-inner product and therefore qualifies as an observable in the quasi-Hermitian framework; the quasi-Hermiticity condition for the Hamiltonian then takes the compact form $H^{\ddagger}=H$.

Let $|\psi_n\rangle$ denote the right eigenstates of a diagonalizable quasi-Hermitian Hamiltonian, $H |\psi_n\rangle = E_n |\psi_n\rangle$ with $E_n\in\mathbb{R}$.  From $H^{\ddagger}=H$ one finds that the left eigenstates can be chosen as $|\phi_n\rangle = \eta |\psi_n\rangle$.  They satisfy the biorthogonality relations $\langle \phi_m | \psi_n \rangle = \delta_{mn}$ and the completeness $\sum_n |\psi_n\rangle \langle \phi_n| = \mathbb{I}$, which guarantee the expansion $|\psi\rangle = \sum_n \langle \phi_n | \psi \rangle\, |\psi_n\rangle$.  In the physical Hilbert space, the right eigenstates themselves form an orthonormal basis, $(\psi_m,\psi_n)_{\eta} = \langle \psi_m | \eta | \psi_n \rangle = \delta_{mn}$, and the spectral decomposition takes the standard form $H = \sum_n E_n\, |\psi_n\rangle \langle \phi_n|$.

We consider here the time-independent case where both $H$ and $\eta$ are constant.  The dynamics is governed by the Schr\"odinger equation $\mi\frac{\dif}{\dif t}|\psi(t)\rangle = H |\psi(t)\rangle$, with the time-evolution operator $U(t) = \me^{-\mi Ht}$.  Using $H^{\ddagger}=H$, it follows that $U(t)$ is $\eta$-unitary,
\begin{align}
U^{\ddagger}(t)U(t) = \me^{\mi H^{\ddagger}t}\me^{-\mi Ht} = \mathbb{I},
\qquad
U^\dagger(t)\,\eta\,U(t) = \eta .
\end{align}
Equivalently, the physical metric $\eta$ is preserved by the time evolution, i.e., it remains invariant under $U(t)$.  Hence, probability is conserved under the physical inner product, $(\psi(t),\psi(t))_{\eta} = (\psi(0),\psi(0))_{\eta}$, and the time-evolved state is simply $|\psi(t)\rangle = \me^{-\mi Ht} |\psi(0)\rangle$.

\subsection{Dynamical evolution of the density matrix in quasi-Hermitian systems}

To extend the pure-state evolution to mixed states, we now formulate the dynamics of a density matrix $\rho$ in the quasi-Hermitian framework.  For a quasi-Hermitian Hamiltonian $H = H^{\ddag}$ with real spectrum, a thermal equilibrium state at inverse temperature $\beta$ is naturally defined by the usual Gibbs form
\begin{align}
\rho = \frac{\me^{-\beta H}}{Z}, \quad
Z = \mathrm{Tr}\bigl(\me^{-\beta H}\bigr) = \sum_n \me^{-\beta E_n}.
\end{align}
Because any analytic function of $H$ inherits $\eta$-self-adjointness, the equilibrium density matrix satisfies $\rho^{\ddag} = \rho$.  Although $\rho$ is non-Hermitian under the standard Dirac inner product, it is a genuine self-adjoint operator in the physical Hilbert space $(\mathcal{H}, (\cdot,\cdot)_\eta)$, thereby admitting a consistent statistical interpretation.

Under a time-independent quasi-Hermitian Hamiltonian $H$, the time evolution of a mixed state is required to preserve $\eta$-self-adjointness and the normalization $\mathrm{Tr}\,\rho = 1$.  This is achieved by the natural extension of pure-state $\eta$-unitary evolution,
\begin{align}
\rho(t) = U(t)\,\rho(0)\,U^{\ddag}(t), \qquad
U(t) = \me^{-\mi H t}.
\end{align}
Differentiation gives the dynamical equation
\begin{align}
\mi \frac{\dif \rho(t)}{\dif t} = H\rho - \rho H^{\ddag},
\end{align}
which, owing to $H^{\ddag}=H$, formally coincides with the von Neumann equation of Hermitian quantum mechanics, albeit with the inner product and geometric structure fixed by $\eta$.  One immediately checks that $\rho^{\ddag}(t)=\rho(t)$ and $\mathrm{Tr}\,\rho(t)=\mathrm{Tr}\,\rho(0)$ follow directly from $U^{\ddag}(t)U(t)=\mathbb{I}$, so both properties are preserved by the dynamics.  In this precise sense, the mixed-state evolution is not an independent postulate but the statistical extension of the $\eta$-unitary pure-state evolution established in Sec.~\ref{Sec2A}.

\subsection{Purification of the density matrix}\label{Sec2C}

The Loschmidt amplitude for a pure state is defined as $\mathcal{G}(t) = \langle \psi(0) | \psi(t) \rangle$.  To generalize this concept to mixed states, it is highly convenient to represent the density matrix as a ``square root'' of a pure state in an enlarged Hilbert space, a procedure known as purification~\cite{Uhlmann86}.  In the quasi-Hermitian setting, the $\eta$-self-adjointness of $\rho$ guarantees that it can be written as $\rho = W W^{\ddag}$,
where $W$ is the purification (or amplitude) operator.  In complete analogy with the Hermitian case, a spectral decomposition reads
\begin{align}
\rho = \sum_n \lambda_n |\psi_n\rangle \langle \phi_n|, \quad
W = \sum_n \sqrt{\lambda_n}\, |\psi_n\rangle \langle \phi_n|\,\mathcal{U},
\end{align}
with $\lambda_n = \me^{-\beta E_n}/Z$ and $|\psi_n\rangle, \langle\phi_n|$ the biorthogonal eigenstates of $H$ introduced in Sec.~\ref{Sec2A}.  The operator $\mathcal{U}$ is an arbitrary element of the quasi-Hermitian unitary group, satisfying $\mathcal{U}^{\ddag}\mathcal{U}=\mathbb{I}$, and reflects a gauge degree of freedom inherent in the purification: the transformation $W \to W\,\mathcal{U}$ leaves the physical density matrix $\rho = W W^{\ddag}$ invariant.

To formulate quench dynamics in a language that directly parallels the pure-state case, it is useful to pass from the operator $W$ to its pure-state representation.  This is achieved by mapping $W$ to a vector in the doubled Hilbert space $\mathcal{H}_S \otimes \mathcal{H}_A$,
\begin{align}
|W\rangle = \sum_n \sqrt{\lambda_n}\, |\psi_n\rangle \otimes \mathcal{U}^T |\phi_n\rangle,
\end{align}
where the first factor belongs to the system and the second to an ancillary space.  The overlap of two purified states is measured by the physical inner product on the doubled space.  Since the physical metric on $\mathcal{H}_S$ is $\eta$, the requirement that the $\eta$-Hilbert-Schmidt inner product $\mathrm{Tr}(W_1^{\ddag} W_2)$ be reproduced as a vector inner product on $\mathcal{H}_S \otimes \mathcal{H}_A$ forces the metric on the ancillary space to be $\eta^{-1}$~\cite{Hou2026UhlmannQuasiHermitian}.  The doubled-space metric is therefore $\eta \otimes \eta^{-1}$, and we denote the resulting inner product by
\begin{align}\label{HS}
( W_1 , W_2 )_{\eta} := \langle W_1 | W_2 \rangle_{\eta\otimes\eta^{-1}} = \mathrm{Tr}(W_1^{\ddag} W_2),
\end{align}
where the equality with the $\eta$-Hilbert-Schmidt trace is verified in Appendix~\ref{purification-inner}.  A crucial point for what follows is that any gauge choice $\mathcal{U}$ cancels out in this overlap, so that physical quantities derived from purified states do not depend on the auxiliary operator $\mathcal{U}$.

\subsection{Theory of metric-stationary quench}

Having developed the requisite tools for mixed-state dynamics and purification in quasi-Hermitian systems, we now apply them to a quantum quench scenario.  Consider a system prepared in a mixed state $\rho(0)$ governed by an initial quasi-Hermitian Hamiltonian $H_i$, satisfying $H_i^\dagger = \eta_i H_i \eta_i^{-1}$ with a positive-definite metric $\eta_i$.  At $t=0$, a parameter of the Hamiltonian is suddenly changed so that the dynamics for $t>0$ is generated by a final quasi-Hermitian Hamiltonian $H_f$, satisfying $H_f^\dagger = \eta_f H_f \eta_f^{-1}$.  In the most general setting, the metric operators $\eta_i$ and $\eta_f$ can differ, leading to a technically involved description because the physical inner product itself depends on time.  To isolate the essential physics of thermally driven dynamical phase transitions, we focus in this work on the important subclass of \emph{metric-stationary quenches}, where both Hamiltonians share the same fixed, parameter-independent metric $\eta$:
\begin{equation}
H_i^\dagger = \eta H_i \eta^{-1}, \qquad
H_f^\dagger = \eta H_f \eta^{-1}.
\end{equation}
This assumption means that the geometry of the physical Hilbert space remains unchanged during the quench; only the generator of time evolution is altered.  Consequently, the $\eta$-adjoint operation $A^{\ddagger} = \eta^{-1}A^\dagger\eta$ is unique and the time evolution of the density matrix is simply
\begin{equation}\label{eq:quench-rho}
\rho(t) = \me^{-\mi H_f t}\,\rho(0)\,\me^{\mi H_f^{\ddag} t},
\end{equation}
which manifestly preserves $\mathrm{Tr}\,\rho(t) = \mathrm{Tr}\,\rho(0)$ because $\me^{\mi H_f^{\ddag} t}\me^{-\mi H_f t}=\mathbb{I}$.

Within the purification framework of Sec.~\ref{Sec2C}, we write $\rho(0)=W(0)W(0)^{\ddag}$.  Since the quench only acts on the system degrees of freedom while leaving the ancillary space unchanged, the time-evolved purification is obtained by $W(t) = \me^{-\mi H_f t}\,W(0)$.
The mixed-state Loschmidt amplitude is then naturally defined as the overlap between the purified states before and after the quench, measured with the physical inner product of the doubled Hilbert space:
\begin{equation}\label{eq:def-LA}
\mathcal{G}(t) = \langle W(0) | W(t) \rangle_{\eta\otimes\eta^{-1}}.
\end{equation}
Substituting the spectral decomposition of the initial thermal state $W(0) = \sum_n \sqrt{\lambda_n^{(i)}} |\psi_n^{(i)}\rangle\langle\phi_n^{(i)}|\,\mathcal{U}$,
and using the biorthogonality relations together with $\mathcal{U}^{\ddag}\mathcal{U}=\mathbb{I}$, one straightforwardly evaluates the overlap:
\begin{align}\label{eq:mixed-LA-final}
\mathcal{G}(t)
&= \sum_{n} \lambda_n^{(i)} \langle\phi_n^{(i)}|\me^{-\mi H_f t}|\psi_n^{(i)}\rangle =\mathrm{Tr}\bigl[\rho(0)\,\me^{-\mi H_f t}\bigr].
\end{align}
Notice that the arbitrary gauge operator $\mathcal{U}$ has dropped out, confirming that the Loschmidt amplitude is an intrinsic physical quantity.  Equation~(\ref{eq:def-LA}) reduces to the standard pure-state result $\mathcal{G}(t) = \langle \psi(0) | \psi(t) \rangle_\eta$ when $\rho(0)$ is a pure state. We note that Eq.~(\ref{eq:mixed-LA-final}) is not an ad hoc definition but follows directly from the first-principles construction developed above, namely, the thermal state, its purification, the gauge structure, and the $\eta$-unitary quench evolution.
To the best of our knowledge, such a systematic and self-contained derivation for quasi-Hermitian systems has not been presented before.

As in Hermitian systems, zeros of $\mathcal{G}(t)$ signal dynamical quantum phase transitions (DQPTs).  The corresponding rate function is introduced in analogy with the free-energy density $r(t) = -\lim_{L\to\infty}\frac{1}{L}\ln\bigl|\mathcal{G}(t)\bigr|^2$,
whose non-analyticities at critical times $t^*$ mark the DQPTs.  The metric-stationary framework ensures that the inner-product structure, the purification, and the Loschmidt amplitude are all defined within a single consistent physical Hilbert space, free from ambiguities arising from parameter-dependent metrics.

\section{Thermal suppression of DQPTs in a multi-level model}\label{Sec3}

With the general formalism established, we now turn to a concrete model to investigate how temperature affects the DQPTs of quasi-Hermitian systems.
In particular, we ask whether tuning the temperature of the initial thermal state can qualitatively alter the dynamical critical behavior, and whether a temperature-driven transition between the presence and the complete absence of DQPTs can occur.

\subsection{General setup: a two-level sector coupled to spectator states}

To make the influence of temperature as pronounced as possible, we need a mechanism by which thermal excitations can significantly redistribute the spectral weights that govern the Loschmidt amplitude.
A minimal strategy is to embed the nontrivial dynamics within a larger Hilbert space that contains additional, inactive energy levels.
These extra levels do not participate in the coherent dynamics of the quench, but their thermal occupation weights compete with those of the active levels.
As temperature increases, the growing Boltzmann weights of these spectator states can dilute the interference signal from the dynamical sector, eventually suppressing the DQPTs altogether.

\subsubsection{Model Hamiltonian and the quench protocol}
Concretely, we consider a general finite-dimensional $\mathcal{PT}$-symmetric Hamiltonian constructed as a direct sum of a nontrivial two-level sector and $N-2$ ($N\ge3$) spectator states:
\begin{align}
H = H_{2} \oplus \bigoplus_{j=1}^{N-2} R^{(j)} .
\end{align}
The two-level block takes the standard form
\begin{align}\label{eq:H2block}
H_{2} = \boldsymbol{R}\cdot\boldsymbol{\sigma}
= (a\,\boldsymbol{n}^r + \mi b\,\boldsymbol{n}^\theta)\cdot \boldsymbol{\sigma},
\end{align}
where $\boldsymbol{\sigma}=(\sigma_x,\sigma_y,\sigma_z)^T$ are Pauli matrices, the unit vector $\boldsymbol{n}^r=
(\sin\theta\cos\phi,\;
 \sin\theta\sin\phi,\;
 \cos\theta)^T$
 denotes the radial direction on the Bloch sphere, and $\boldsymbol{n}^\theta
=
(\cos\theta\cos\phi,\;
 \cos\theta\sin\phi,\;
 -\sin\theta)^T$
is one of the tangent directions at the point $(\theta,\phi)$~\cite{Wang_2010}.
The eigenvalues of $H_{2}$ are $E_\pm =\pm \sqrt{a^2-b^2}$ in the unbroken $\mathcal{PT}$ regime $a^2>b^2$.
The spectator energies $R^{(j)}$ are real constants, so the full Hamiltonian is a direct sum of quasi-Hermitian blocks and remains quasi-Hermitian with respect to a block-diagonal metric that will be specified in the examples below.

The initial state is the thermal equilibrium state of the pre-quench Hamiltonian $H_i$, i.e., $\rho(0) = \me^{-\beta H_i}/Z_i$ with $Z_i = \mathrm{Tr}[\me^{-\beta H_i}]$.
In the two-level sector the eigenenergies are $\pm R_i = \pm\sqrt{a_i^2 - b_i^2}$; for the spectator sector we assume, for simplicity, that all initial eigenenergies are identical to the excited-state energy of the two-level sector, $R^{(1)}_i = R^{(2)}_i = \cdots = R^{(N-2)}_i = R_i$.
A quench is performed by suddenly changing the parameters of the two-level block and, optionally, the spectator energies.  For convenience we also set all post-quench spectator energies to a constant: $R^{(1)}_f = \cdots = R^{(N-2)}_f = R$.
For other parameters, the quench results in $(a_i,b_i,\theta_i,\phi_i)\rightarrow (a_f,b_f,\theta_f,\phi_f)$.
To remain within the metric-stationary framework, we must impose that the pre- and post-quench Hamiltonians share the same positive-definite metric.
For the $\mathcal{PT}$-symmetric two-level Hamiltonian, the metric $\eta_2$ is known to depend on the ratio $b/a$ and the azimuthal angle $\phi$ as~\cite{Wang_2010}
\begin{align}
\eta_2(\alpha,\phi)=\frac{1}{\cos\alpha}
\begin{pmatrix}
1 & \mi\sin\alpha\,\me^{-\mi\phi} \\
-\mi\sin\alpha\,\me^{\mi\phi} & 1
\end{pmatrix},
\end{align}
where $\sin\alpha = \frac{b}{a}$.
Hence, $\eta_i = \eta_f$ if and only if
\begin{align}
\alpha_i = \alpha_f \;\Longleftrightarrow\; \frac{b_i}{a_i} = \frac{b_f}{a_f},
\quad\text{and}\quad
\phi_i = \phi_f.
\end{align}
In the Hermitian limit $b_i=b_f=0$, the metric reduces to the identity, $\eta_i=\eta_f=1_{2\times 2}$, and the formalism recovers the standard quench dynamics of a Hermitian quantum system.

\subsubsection{Loschmidt amplitude}
To evaluate the Loschmidt amplitude, the block-diagonal structure of both $H_i$ and $H_f$ yields the factorizations $\me^{-\beta H_i} = \me^{-\beta H_{2i}} \oplus \bigoplus_{j=1}^{N-2}\me^{-\beta R_i}$ and $\me^{-\mi H_f t} = \me^{-\mi H_{2f} t} \oplus \bigoplus_{j=1}^{N-2} \me^{-\mi Rt}$.
The trace therefore separates into a dynamical two-level contribution and a static spectator contribution.
Within the two-level sector one has
\begin{align}
\me^{-\beta H_{2i}} &= \cosh(\beta R_{i}) - \sinh(\beta R_{i})\,\hat{\boldsymbol{R}}_{i}\cdot\boldsymbol{\sigma}, \notag\\
\me^{-\mi H_{2f} t} &= \cos(R_f t) - \mi\sin(R_f t)\,\hat{\boldsymbol{R}}_f\cdot\boldsymbol{\sigma},
\end{align}
where $\hat{\boldsymbol{R}}_{i,f}=\boldsymbol{R}_{i,f}/R_{i,f}$.
Substituting these expressions into Eq.~(\ref{eq:mixed-LA-final}) and evaluating the trace over the block-diagonal structure, we obtain
\begin{align}\label{eq:GN}
\mathcal{G}(t)&=\frac{\text{Tr}[\me^{-\beta H_{2i}} \me^{-\mi H_{2f} t}] + (N-2)\me^{-\beta R_i}\me^{-\mi Rt}}{\text{Tr}[\me^{-\beta H_{2i}}] + (N-2)\me^{-\beta R_i}}\notag\\
&=\frac{
  \left\{
    \begin{aligned}
     & 2\cos(R_f t)\cosh(\beta R_i) \\
    +& 2\mi\sin(R_f t)\sinh(\beta R_i)\hat{\boldsymbol{R}}_i\cdot\hat{\boldsymbol{R}}_f \\
    +& (N-2)\me^{-\beta R_i}\me^{-\mi Rt}
    \end{aligned}
  \right\}
}{
  2\cosh(\beta R_i)+(N-2)\me^{-\beta R_i}
}.
\end{align}
Here $\hat{\boldsymbol{R}}_i\cdot\hat{\boldsymbol{R}}_f = u + \mi v$
where
\begin{align}
u &= \frac{1}{R_i R_f}\Bigl[
      a_i a_f\bigl(\sin\theta_i\sin\theta_f\cos\Delta\phi + \cos\theta_i\cos\theta_f\bigr) \notag\\
      &\qquad\quad - b_i b_f\bigl(\cos\theta_i\cos\theta_f\cos\Delta\phi + \sin\theta_i\sin\theta_f\bigr)
      \Bigr],\notag\\
v &= \frac{a_i b_f + a_f b_i}{2R_i R_f}\,\sin(\theta_i+\theta_f)(\cos\Delta\phi - 1),
\end{align}
with $R_{i,f} = \sqrt{a_{i,f}^2 - b_{i,f}^2}$ and $\Delta\phi \equiv \phi_i - \phi_f$.
In the Hermitian limit $b_i=b_f=0$, the imaginary part vanishes and $\hat{\boldsymbol{R}}_i\cdot\hat{\boldsymbol{R}}_f$ reduces to the known real-valued result~\cite{PhysRevB.111.174310} (assuming $a_{i,f}>0$ without loss of generality):
\begin{align}
\hat{\boldsymbol{R}}_i\cdot\hat{\boldsymbol{R}}_f=\cos\theta_i\cos\theta_f+\sin\theta_i\sin\theta_f\cos(\phi_i-\phi_f).
\end{align}
The metric-stationary condition $\Delta\phi=0$ likewise forces $v=0$, while the real part simplifies to
\begin{align}
u =\frac{a_i a_f - b_i b_f}{R_i R_f}\cos(\theta_i-\theta_f).
\end{align}
This condition also requires $\sin\alpha = b_i/a_i = b_f/a_f$, from which one finds $a_i a_f - b_i b_f = a_i a_f\cos^2\alpha$ and $R_i R_f = |a_i a_f|\cos^2\alpha$.  Consequently,
\begin{align}
u =\text{sgn}(a_i a_f)\cos(\theta_i-\theta_f)=\cos(\theta_i-\theta_f)
\end{align}
under the assumption $a_{i,f}>0$.

\subsubsection{Properties of DQPTs}

Even the simple quench protocol described above can produce interesting phenomena.  We consider two classes of quenches.

\emph{Case 1: $R=0$ after quench.}
In this case Eq.~(\ref{eq:GN}) clearly exposes the interplay between the dynamical two-level sector and the spectator states: the first two terms in the numerator arise from the coherent interference of the two active levels, while the last term is the purely static contribution of the $N-2$ spectators.
As $N$ increases or as the temperature grows, the relative weight of the spectator term increases, progressively diluting the interference signal.

According to Eq.~(\ref{eq:GN}), the DQPT condition $\mathcal{G}(t^*_n)=0$ requires both the real and imaginary parts of the numerator to vanish.  The imaginary part vanishes when $\hat{\boldsymbol{R}}_i\cdot\hat{\boldsymbol{R}}_f = 0$, and the real part then yields
\begin{align}\label{t*n1}
\cos(R_f t^*_n) = -\frac{N-2}{1+\me^{2\beta R_i}}.
\end{align}
For real-valued critical times to exist, the right-hand side must lie within $[-1,1]$, which leads to the existence condition $\me^{2\beta R_i} \ge N-3 $.
Thus a threshold temperature emerges:
\begin{align}
T_{\rm th}(N) = \frac{2R_i}{\ln(N-3)}.
\end{align}
For $N=2$, the condition $\me^{2\beta R_i}\ge -1$ is trivially satisfied and DQPTs persist at all temperatures.
For $N=3$, the condition becomes $\me^{2\beta R_i}\ge 0$, which holds for all finite $\beta$; DQPTs are never completely suppressed, but their critical times acquire a nontrivial temperature dependence.
For $N=4$, the condition requires $\me^{2\beta R_i}\ge 1$, meaning the threshold temperature formally diverges; suppression occurs only in the extreme high-temperature limit.
For $N\geq 5$, a finite, physically accessible threshold temperature $T_{\rm th}(N)$ emerges, above which DQPTs disappear.

Equation~(\ref{eq:GN}) defines a critical curve in the $(t,T)$ plane.  For a given temperature below $T_{\rm th}$ it determines the critical times $t_n^*$; conversely, for a fixed instant satisfying $\cos(R_f t)<0$, one may solve for the temperature $T^*$ at which that particular instant becomes critical:
\begin{align}
T^*(t) = \frac{2R_i}{\ln\!\bigl(-\frac{N-2}{\cos(R_f t)}-1\bigr)}.
\end{align}
This is merely a reparameterization of the same critical condition and does not constitute a temperature-driven DQPT, since the nonanalyticity always resides in the time domain.  The physically meaningful role of temperature is captured by $T_{\rm th}$: it sets the upper bound of the temperature window within which DQPTs exist.  For this reason we refer to $T_{\rm th}$ as a threshold temperature rather than a critical temperature, to distinguish it from the conventional critical points of equilibrium phase transitions.

%We now illustrate these behaviors with explicit examples.

\emph{Case 2: $R=R_f$ after quench.}
To introduce a stronger non-Hermitian signature, we now set the post-quench spectator energies equal to the two-level gap $R_f$, so that the spectator sector is no longer static but dynamically modulated by the same energy scale that governs the two-level block.
Equation~(\ref{eq:GN}) then contains an additional oscillatory phase in the last term.
Under the metric-stationary condition $\Delta\phi=0$, the scalar product $\hat{\boldsymbol{R}}_i\cdot\hat{\boldsymbol{R}}_f = u$ is real, and the numerator can be reorganized into real and imaginary parts.
A straightforward calculation yields
\begin{align}\label{eq:G_Rf}
\mathcal{G}(t) &= \cos(R_f t) \notag\\&+ \mi \sin(R_f t)\,\frac{2\sinh(\beta R_i)\,u - (N-2)\me^{-\beta R_i}}{2\cosh(\beta R_i) + (N-2)\me^{-\beta R_i}}.
\end{align}
The DQPT condition $\mathcal{G}(t^*_n)=0$ requires the real and imaginary parts of the numerator to vanish simultaneously.
The real part vanishes when $\cos(R_f t)=0$, i.e., $R_f t_n^* = (n+1/2)\pi$, at which $\sin(R_f t_n^*)=\pm1\neq0$.
The vanishing of the imaginary part therefore imposes
\begin{align}
2\sinh(\beta R_i)\,u - (N-2)\me^{-\beta R_i} = 0 \Longrightarrow u = \frac{N-2}{\me^{2\beta R_i} - 1}.
\end{align}
Since the right-hand side is positive for all finite temperatures when $N>2$, a DQPT requires $u>0$, i.e., $|\theta_i-\theta_f|<\pi/2$.
Combined with the metric-stationary bound $u \le 1$, a real solution exists only when $\frac{N-2}{\me^{2\beta R_i} - 1} \le 1 \Longleftrightarrow \me^{2\beta R_i} \ge N-1$.
This yields the threshold temperature
\begin{align}\label{eq:Tth_Rf}
T_{\rm th}(N) = \frac{2R_i}{\ln(N-1)}.
\end{align}
This result differs qualitatively from the $R=0$ case, where $T_{\rm th}(N) = 2R_i/\ln(N-3)$ and a finite threshold appears only for $N\ge5$.

A more striking difference is the structure of the DQPT condition itself.  In the $R=R_f$ protocol, DQPTs are restricted to the discrete time slices $t_n^* = (n+1/2)\pi/R_f$ by the vanishing of the real part.  On each such slice, the imaginary part selects a unique temperature
\begin{align}\label{eq:Tstar_Rf}
T^* = \frac{2R_i}{\ln\!\bigl(1+\frac{N-2}{u}\bigr)}.
\end{align}
Comparing $T^*$ with $T_{\rm th}$ gives $T^* \le T_{\rm th}$, with equality only for the trivial limit $u=1$ ($\theta_i=\theta_f$); thus $T^*$ always lies within the window where DQPTs are permitted.
This is fundamentally different from the $R=0$ case, where a continuous temperature window below $T_{\rm th}$ supports DQPTs.  Here the time dimension is rigid: DQPTs can only occur at the prescribed instants $t_n^*$.  The temperature dimension is soft: on each such slice one may tune $T$ to $T^*$ to hit the zero.  This time-temperature asymmetry reflects the defining feature of DQPTs as time-driven phenomena, with temperature acting as a control parameter that selects whether and where the critical times occur.

\subsubsection{Effects of non-Hermitian parameters}
The metric-stationary condition $\eta_i=\eta_f$ forces $b_i/a_i=b_f/a_f$ and $\phi_i=\phi_f$, reducing the overlap parameter to $\hat{\boldsymbol{R}}_i\cdot\hat{\boldsymbol{R}}_f = \cos(\theta_i-\theta_f)\in[-1,1]$,
independently of $b_{i,f}$.  The imaginary part $v$ therefore vanishes identically, and no complex interference channel appears.
The non-Hermitian parameters enter solely through the renormalized two-level gaps $R_{i,f}=\sqrt{a_{i,f}^2-b_{i,f}^2}=a_{i,f}\cos\alpha$,
with $\sin\alpha=b_{i,f}/a_{i,f}$ fixed by the metric-stationary condition.
This renormalization shifts the critical temperature and the critical times quantitatively.  In the $R=R_f$ protocol, the spectator oscillations compete with the two-level interference, giving rise to a temperature-selective DQPT that is absent in the static-spectator case.  The Hermitian limit $b_{i,f}\to0$ is recovered smoothly throughout.

The absence of genuinely non-Hermitian dynamical signatures is a direct consequence of the equivalence theorem that will be established in Sec.~\ref{Sec5}: under a fixed positive-definite metric, the quasi-Hermitian Loschmidt amplitude is identical to that of the associated Hermitian system $h=\eta^{1/2}H\eta^{-1/2}$.  The parameters $b_{i,f}$ are absorbed into the similarity transformation and manifest themselves only as a renormalization of the Hermitian energy scale.  Complex interference terms or parameter-space holonomies require relaxing the fixed-metric assumption, as will be discussed in Sec.~\ref{Sec5}.

% ============================================================
\subsection{Numerical analysis}
\subsubsection{Case 1: $R=0$ after quench}

To make the influence of temperature as pronounced as possible, we embed the nontrivial dynamics within a larger Hilbert space that contains additional, inactive energy levels.  These extra levels do not participate in the coherent dynamics of the quench, but their thermal occupation weights compete with those of the active levels.  As temperature increases, the growing Boltzmann weights of these spectator states can dilute the interference signal from the dynamical sector, eventually suppressing the DQPTs altogether.

For all examples below we adopt the same two-level parameters: $a_i=a_f=3$, $b_i=b_f=\sqrt{8}$, $\theta_i=0$, $\theta_f=\pi/2$, and $\phi_i=\phi_f=0$.  With this choice the metric-stationary condition $\eta_i=\eta_f$ is satisfied because $b_i/a_i=b_f/a_f$ and $\phi_i=\phi_f$, so the physical inner product remains unchanged across the quench.  The quench alters the polar angle $\theta$ and sets the post-quench spectator energies to $R=0$.
In this case the positive-definite metric takes the block-diagonal form $\eta = \eta_2 \oplus 1_{N-2}$, where
\begin{align}\label{eta2}
\eta_2 =
\begin{pmatrix}
3 & 2\mi\sqrt{2}  \\
-2\mi\sqrt{2} & 3
\end{pmatrix},
\end{align}
and $1_{N-2}$ denotes the $(N-2)\times (N-2)$ identity matrix.  This block-diagonal structure ensures that $H_{i,f}^\ddagger = H_{i,f}$ for any $N$.

Using the general $N$-level expression (\ref{eq:GN}) with $\hat{\boldsymbol{R}}_i\cdot\hat{\boldsymbol{R}}_f=0$ and $R_i=R_f=1$, the Loschmidt amplitude simplifies to
\begin{align}\label{eq:LA_N}
\mathcal{G}(t) = \frac{\cos(t)\bigl(1+\me^{2\beta}\bigr) + N-2}{N-1+\me^{2\beta}} .
\end{align}
The DQPTs occur when $\mathcal{G}(t)$ vanishes, which yields the unified critical condition
\begin{align}\label{eq:unified_critical}
\cos(t^{*(N)}_n) = -\frac{N-2}{1+\me^{2\beta}} ,
\end{align}
and the corresponding critical times
\begin{align}\label{eq:unified_tn}
t^{*(N)}_n = \arccos\!\left(-\frac{N-2}{1+\me^{2\beta}}\right) + 2n\pi, \qquad n\in\mathbb{Z}.
\end{align}
Real-valued solutions exist only when the right-hand side of Eq.~(\ref{eq:unified_critical}) lies within $[-1,1]$, which requires $\me^{2\beta} \ge N-3 $.
This gives the dimensionality-dependent threshold temperature
\begin{align}\label{eq:Tth_N}
T_{\rm th}(N) = \frac{2}{\ln(N-3)} ,
\end{align}
valid for $N\ge 5$.  For $N=3$ the condition is trivially satisfied at all temperatures, while for $N=4$ it requires $\me^{2\beta}\ge 1$ and the threshold temperature formally diverges.

\begin{figure}[ht]
	\centering
	\includegraphics[width=3.4in,clip]{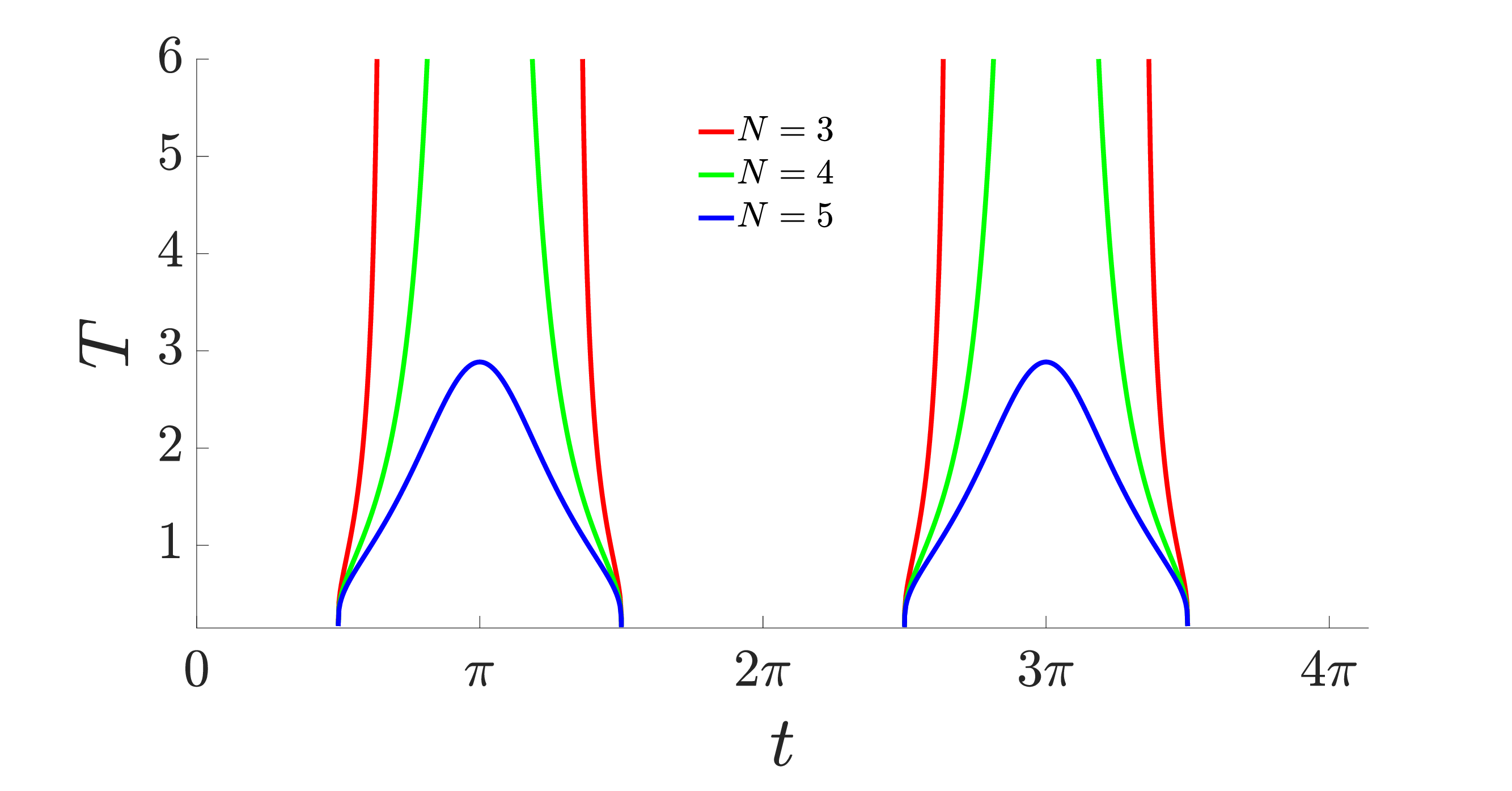}
    \caption{DQPT critical curves in the $(t,T)$ plane for $N=3$ (red), $N=4$ (green), and $N=5$ (blue) under the $R=0$ quench protocol.  For $N=5$ a finite threshold temperature $T_{\rm th}=2/\ln 2$ emerges, above which no real zeros exist and DQPTs are fully suppressed.}
	\label{Fig1}
\end{figure}

Figure~\ref{Fig1} displays the DQPT critical curves in the $(t,T)$ plane for $N=3$, $4$, and $5$.  All curves are periodic with period $2\pi$ and emanate from $t=\pi/2$ (mod $2\pi$) at $T=0$.  For $N=3$ (red), the critical time drifts from $\pi/2$ toward $2\pi/3$ as $T\to\infty$, while for $N=4$ (green) it approaches $\pi$ in the same limit; in both cases DQPTs persist at all finite temperatures.  For $N=5$ (blue), the curve forms a closed loop: starting from $t=\pi/2$ at $T=0$, it rises to a maximum temperature $T_{\rm th}=2/\ln 2$ at $t=\pi$, then descends to $t=3\pi/2$ at $T=0$.  Above $T_{\rm th}$ the inequality fails, the Loschmidt amplitude has no real zeros, and the rate function $r(t)$ becomes analytic for all $t$.

The suppression mechanism is traced to the thermal redistribution of spectral weights in the biorthogonal decomposition of the initial state.  At low temperatures, the thermal state is dominated by low-lying eigenmodes, preserving sufficient phase coherence to produce Fisher zeros.  As temperature increases, the Boltzmann weights of the spectator states grow, progressively diluting the relative amplitude of the coherent two-level interference term.  Once the temperature exceeds $T_{\rm th}$, the spectator background overwhelms the interference contribution, and the Loschmidt amplitude can no longer vanish.  This mechanism becomes progressively more effective as the Hilbert-space dimension increases, suggesting that higher-dimensional non-Hermitian systems are generically more susceptible to thermal suppression of dynamical criticality.

\begin{figure}[ht]
	\centering
	\includegraphics[width=3.4in,clip]{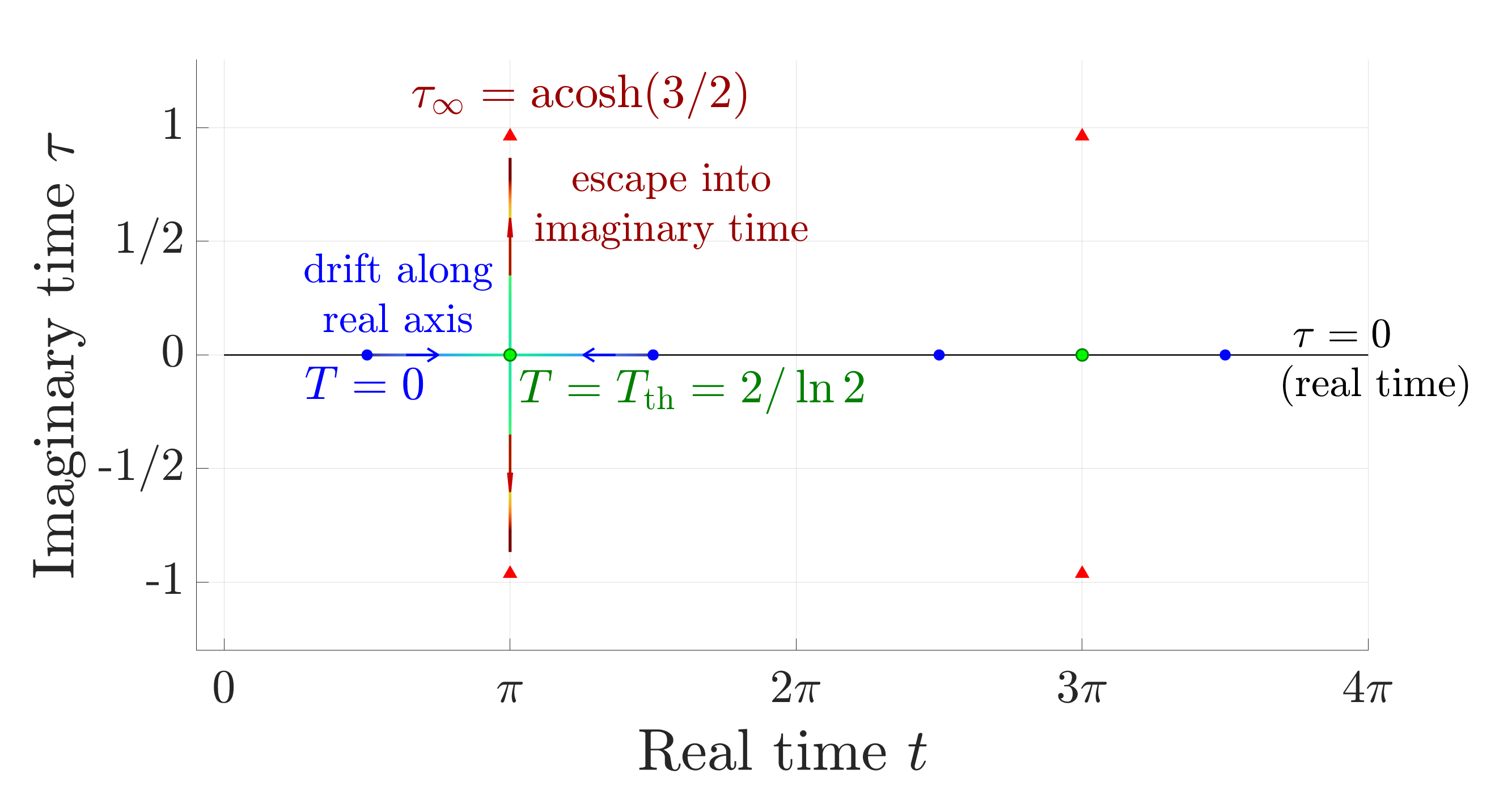}
    \caption{Fisher zeros of the Loschmidt amplitude in the complex time plane for Case~1 ($N=5$, $R=0$).  Blue circles mark the $T=0$ zeros, green circles mark the merger at $T=T_{\rm th}$, and red triangles mark the asymptotic positions as $T\to\infty$.}
	\label{Fig1b}
\end{figure}

\textit{Complex-time picture.}
The threshold temperature $T_{\rm th}$ admits a revealing geometric interpretation in terms of the Fisher zeros of the Loschmidt amplitude.
Extending $\mathcal{G}(t)$ to the complex time plane $z = t + \mi\tau$, and using Eq.~(\ref{t*n1}) with $R_f=1$, the DQPT condition $\mathcal{G}(z)=0$ becomes
\begin{align}
\cos(t + \mi\tau) = -\frac{N-2}{1+\me^{2\beta}} .
\end{align}
Expanding the left-hand side as $\cos(t+\mi\tau) = \cos t\cosh\tau - \mi\sin t\sinh\tau$ and noting that the right-hand side is real, we obtain $\sin t \sinh\tau = 0$
and $\cos t \cosh\tau = -\frac{N-2}{1+\me^{2\beta}}$. 
The first equation yields two solution branches: $\tau = 0$ or $\sin t = 0$.
For $\tau = 0$, the second equation reduces to $\cos t = -\frac{N-2}{1+\me^{2\beta}}$, which has real solutions only when $T \le T_{\rm th}$; this is the real-time DQPT branch.
 For $\sin t = 0$, we have $t = n\pi$ and $\cos t = (-1)^n$.  Since $\cosh\tau \ge 1$ and the right-hand side is negative, $n$ must be odd.  Setting $n = 2m+1$ gives $\cosh\tau = \frac{N-2}{1+\me^{2\beta}}$, which requires $T \ge T_{\rm th}$.  The solutions are $z = (2m+1)\pi \pm \mi\tau(T)$ with $\tau(T) = \operatorname{arccosh}\!\bigl(\frac{N-2}{1+\me^{2/T}}\bigr)$.

The two branches meet at $T = T_{\rm th}$, where $\tau=0$ and $\cos t = -1$, i.e.\ $t = (2m+1)\pi$.  Hence $T_{\rm th}$ marks the coalescence of the real zeros and their subsequent departure into the complex plane along the vertical lines $t = (2m+1)\pi \pm \mi\tau(T)$.  Below $T_{\rm th}$ there are two real zeros per period; above $T_{\rm th}$ the rate function becomes analytic.

Figure~\ref{Fig1b} shows this motion for $N=5$.
Blue circles mark the $T=0$ zeros at $t=\pi/2$ and $3\pi/2$, green circles mark their coalescence at $t=\pi$ and $3\pi$ when $T=T_{\rm th}$, and red triangles indicate the asymptotic positions $\tau=\pm\operatorname{arccosh}(3/2)$ as $T\to\infty$.
This escape of the Fisher zeros from the real axis captures the thermal suppression mechanism in a single geometric image: the zeros drift inward, coalesce at $T_{\rm th}$, and retreat into the imaginary-time direction, leaving no real zeros and a fully analytic rate function.

\subsubsection{Case 2: $R=R_f$ after quench}

We now set the post-quench spectator energies to $R=R_f$, so that the spectator sector oscillates at the same frequency as the two-level block.
The two-level parameters remain $a_i=a_f/2=3$, $b_i=b_f/2=\sqrt{8}$, $\theta_i=0$, $\theta_f=\frac{\pi}{4}$, and $\phi_i=\phi_f=0$, guaranteeing $\eta_i=\eta_f$.
%The quench changes the polar angle $\theta_f$ and leaves the spectator energies unchanged.

With $v=0$ and $R_i=R_f/2=1$, the Loschmidt amplitude reduces to Eq.~(\ref{eq:G_Rf}) with $u=\cos(\theta_i-\theta_f)=\cos\theta_f$.
The DQPT condition forces $\cos t=0$, hence the critical times are the discrete slices $t^*_n = \Bigl(n+\frac12\Bigr)\pi,\quad n\in\mathbb{Z}$,
independently of $N$ and $T$.  At these instants the imaginary part requires $u = \frac{N-2}{\me^{2\beta}-1}$,
which selects a unique temperature $T^* = \frac{2}{\ln\!\bigl(1+\frac{N-2}{u}\bigr)}$ according to Eq.~(\ref{eq:Tstar_Rf}).
For $u=0$ (e.g.\ $\theta_f=\pi/2$) this condition cannot be satisfied when $N>2$, and no DQPT occurs.
To obtain nontrivial behavior we choose $u>0$.

\begin{figure}[ht]
	\centering
	\includegraphics[width=3.4in,clip]{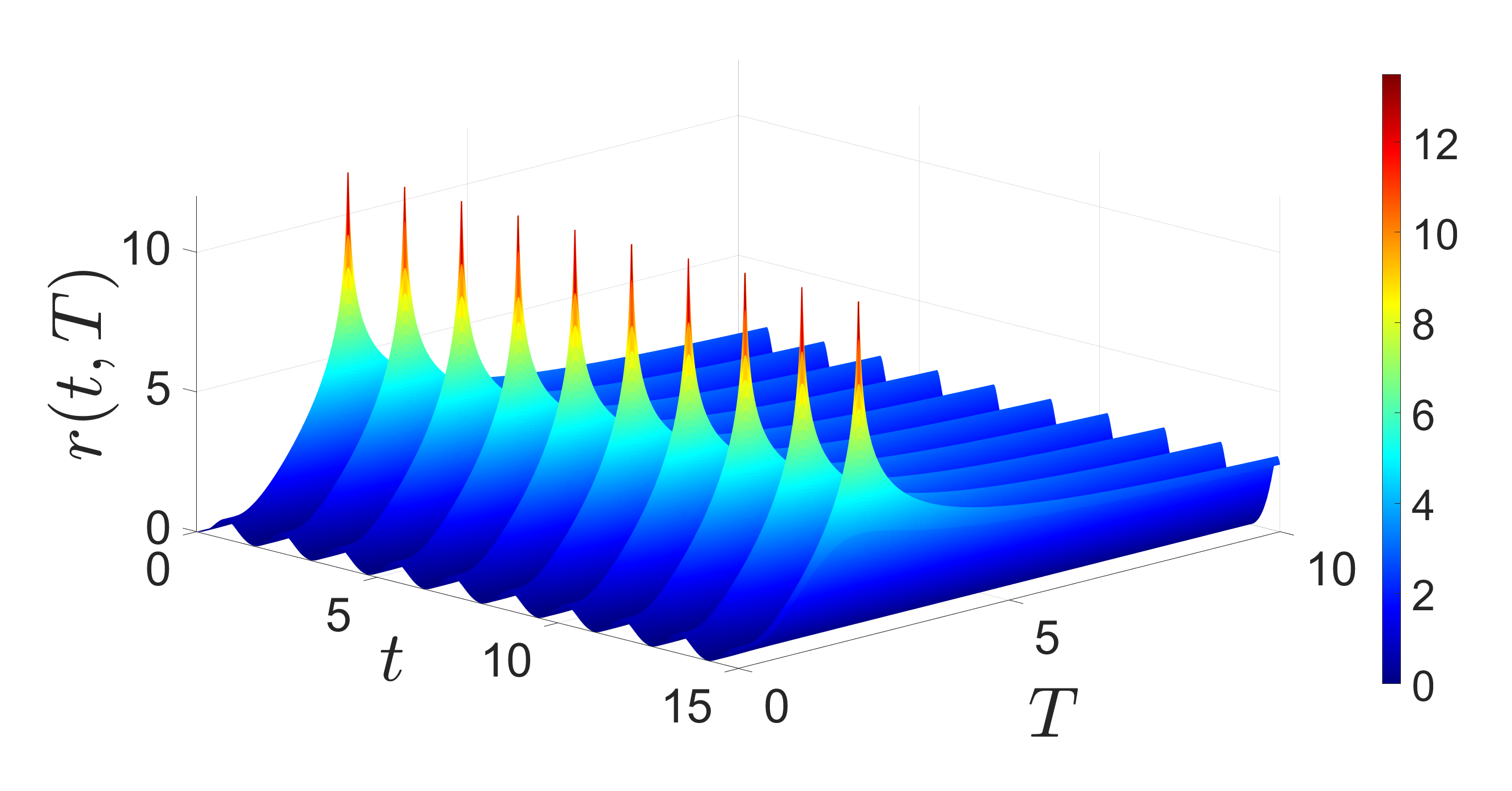}
    \caption{Rate function $r(t,T)$ for the $R=R_f$ quench with $N=3$ and $u>0$.
      Sharp cusps appear exactly at $t^*_n=(n+1/2)\pi$, signaling DQPTs.
      The cusp height decreases with temperature and vanishes above $T^*$, illustrating the temperature-selective suppression of DQPTs.}
	\label{Fig_case2}
\end{figure}

Figure~\ref{Fig_case2} shows $r(t,T)$ for $N=3$.
Pronounced cusps emerge at the prescribed instants $t^*_n$, confirming the DQPTs.
As temperature increases, the growing thermal weight of the spectator states progressively dilutes the interference signal, and the cusps become shallower.
Above $T^*$ the rate function is completely smooth, indicating that the Loschmidt amplitude no longer vanishes.
This temperature-selective suppression is a direct consequence of the competition between spectator oscillations and two-level coherence, and has no analogue in the static-spectator Case~1.

\begin{figure}[ht]
	\centering
	\includegraphics[width=3.4in,clip]{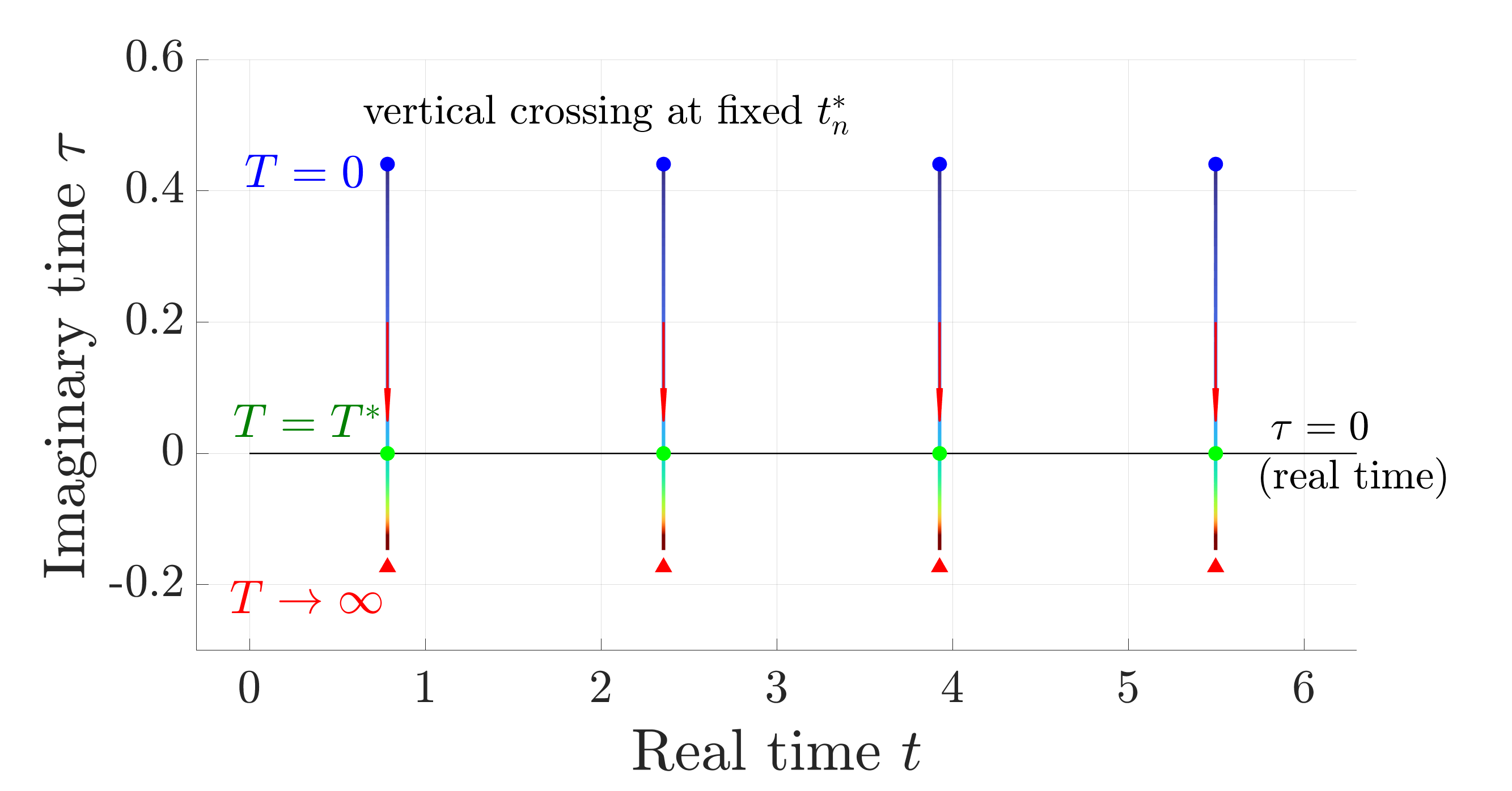}
    \caption{Fisher zeros of the Loschmidt amplitude in the complex time plane for Case~2 ($N=3$, $R=R_f$).  Blue circles mark the $T=0$ positions, green circles mark the real-axis crossing at $T=T^*$, and red triangles mark the asymptotic positions as $T\to\infty$.}
	\label{Fig2b}
\end{figure}

\textit{Complex-time picture.}
Extending the Loschmidt amplitude to the complex time plane sharpens the contrast with Case~1.
For Case~2, with $R_f=2$, $R_i=1$, $N=3$, and $u = 1/\sqrt{2}$, Eq.~(\ref{eq:G_Rf}) gives $\cot(2z) = -\mi\,C(\beta)$, where $C(\beta) \equiv
\frac{\sqrt{2}\sinh\beta - \me^{-\beta}}{2\cosh\beta + \me^{-\beta}}$. 
Using the identity $\cot(x+\mi y) = \frac{\sin(2x) - \mi\sinh(2y)}{\cosh(2y) - \cos(2x)}$, the condition becomes
\begin{align}
\frac{\sin(4t) - \mi\sinh(4\tau)}{\cosh(4\tau) - \cos(4t)} = -\mi\,C(\beta) .
\end{align}
Separating real and imaginary parts gives $\sin(4t)=0$, hence $t = n\pi/4$ with $n\in\mathbb{Z}$.
Only odd $n$ are admissible since $\cos(2t)\neq0$ for even $n$; writing $n = 2m+1$, the critical times are $t_m^* = (2m+1)\pi/4$.
At these instants $\cos(4t_m^*) = -1$, and the imaginary part reduces to $\sinh(4\tau) = C(\beta)(\cosh(4\tau)+1)$, which simplifies to $\tanh(2\tau) = C(\beta)$.
Since $|C(\beta)| < 1$ for all finite temperatures, a real solution $\tau = \frac{1}{2}\operatorname{arctanh}C(\beta)$ always exists.
At $T = T^*$, $C(\beta)=0$ and the zero lies on the real axis ($\tau=0$), giving a physical DQPT.
For $T \neq T^*$ the zero is displaced vertically along $t = t_m^*$, moving from $\tau > 0$ at low temperatures to $\tau < 0$ at high temperatures, and crossing the real axis exactly once.
Note that $T_{\rm th}$ plays a different role here than in Case~1: it does not mark a merger or escape of zeros, but rather sets the theoretical upper bound $T^* \le T_{\rm th}$ for the temperature at which the real-axis crossing can occur.
When $T$ exceeds $T_{\rm th}$, the existence condition $\me^{2\beta R_i} \ge N-1$ is violated and the zero, though still present in the complex plane, can no longer reach the real axis for any $u \in [-1,1]$.

Figure~\ref{Fig2b} shows this motion for $N=3$.
At $T=0$ the zeros sit above the real axis (blue circles); they cross the real axis at $T=T^*$ (green circles); and as $T\to\infty$ they asymptote to $\tau=-\frac{1}{2}\operatorname{arctanh}(1/3)$ (red triangles).
Throughout this evolution the zeros remain pinned to the discrete times $t_m^*=(2m+1)\pi/4$.

This behaviour is fundamentally different from Case~1, where zeros drift along the real axis, coalesce, and escape into the imaginary-time direction.
Here the real part rigidly locks the critical times, while temperature controls only the imaginary displacement.
The single real-axis crossing at $T = T^*$ gives a geometric meaning to the temperature-selective nature of DQPTs in this protocol: among the continuous family of complex zeros, only one lies on the real-time axis and is physically realized.
% ============================================================

% ============================================================
\section{Role of non-Hermiticity under the metric-stationary condition}
\label{Sec4}

The preceding sections have demonstrated that the thermal suppression of DQPTs is controlled by the Hilbert-space dimension and temperature.  A natural question is whether this mechanism is intrinsically non-Hermitian.  At first sight, the equivalence theorem established below might seem to render the quasi-Hermitian formalism superfluous.  However, this conclusion is only valid because the formalism has been consistently developed: without the metric-induced inner product and the associated purification, the very definition of the mixed-state Loschmidt amplitude in a biorthogonal system would be ambiguous.  We now clarify the precise role played by the non-Hermitian parameters, showing that their absorption into a similarity transformation is not a deficiency of the framework but a confirmation of its internal consistency.

A fixed positive-definite metric $\eta$ maps the quasi-Hermitian Hamiltonians to Hermitian ones via the similarity transformation $h_{i,f} \equiv \eta^{1/2} H_{i,f} \eta^{-1/2}$
with $h_{i,f}^\dagger = h_{i,f}$.  Because the metric is unchanged across the quench, the entire protocol admits an equivalent Hermitian representation.  The initial density matrix transforms as $\rho_{\rm H}(0) = \eta^{1/2} \rho(0) \eta^{-1/2}$, and the Loschmidt amplitude becomes
\begin{align}
\mathcal{G}_{\rm NH}(t) \equiv \mathrm{Tr}\bigl[\rho(0)\, \me^{-\mi H_f t}\bigr]
= \mathrm{Tr}\bigl[\rho_{\rm H}(0)\, \me^{-\mi h_f t}\bigr] \equiv \mathcal{G}_{\rm H}(t),
\end{align}
exactly, by the cyclic property of the trace.

Two points must be emphasized.  First, the Hermitian system $h_{i,f}$ is \emph{not} the original Hamiltonian with $b_{i,f}$ set to zero; it is the \emph{associated} Hermitian system obtained via $\eta^{1/2}$, and its spectrum and eigenstates implicitly encode all non-Hermitian parameters.  Second, the equality $\mathcal{G}_{\rm NH}(t) = \mathcal{G}_{\rm H}(t)$ holds pointwise for this specific associated Hermitian system, not for an arbitrary one.

These observations constitute a no-go theorem for metric-stationary quenches: if $H_{i,f}$ share a common positive-definite metric $\eta$, then the Loschmidt amplitude coincides with that of the associated Hermitian system for all $t$.  Consequently, all observables constructed from $\mathcal{G}(t)$, including Fisher zeros, the rate function, and the threshold temperature $T_{\rm th}$, are identical in the two descriptions, and no genuinely non-Hermitian dynamical signatures can arise.

This strict equivalence has two further implications.  First, the thermal suppression mechanism and the emergence of $T_{\rm th}$ for $N\ge 5$ are properties of the multi-level structure, not unique to non-Hermitian physics.  Second, exceptional points are strictly excluded under a fixed positive-definite metric.  This is seen from the threshold temperature $T_{\rm th}(N)=\frac{2\sqrt{a_i^2-b_i^2}}{\ln(N-3)}$,
which depends on the non-Hermitian parameters only through $R_i=\sqrt{a_i^2-b_i^2}$.  As $b_i\to a_i$, the gap closes continuously, $T_{\rm th}\to 0$, and DQPTs are suppressed at any finite temperature.  This is an ordinary spectral degeneracy smoothly connected to the Hermitian limit $b_i=0$, without any non-analyticity or change of algebraic structure.

The absence of non-Hermitian signatures can also be understood from the geometric perspective of Hermitianization~\cite{Wang2026Obstructions}.  A global similarity transformation $S=\eta^{1/2}$ that maps the quasi-Hermitian system to a Hermitian one can fail globally if the parameter space exhibits curvature or nontrivial holonomies.  In the present quench protocol, the path is a straight line from $\theta_i$ to $\theta_f$ at fixed $\phi$, and the metric is independent of $\theta$; hence the connection $\mathcal{G}_\mu=\frac{1}{2}[\partial_\mu\sqrt{\eta},\sqrt{\eta}^{-1}]$ vanishes identically.  No geometric or topological obstruction arises, guaranteeing the strict equivalence.

Within the metric-stationary framework non-Hermiticity is therefore absorbed into a global similarity transformation and manifests only as a renormalization of energy scales.  The thermal suppression of DQPTs and the finite threshold temperature $T_{\rm th}$ are universal features of the multi-level structure, not unique signatures of non-Hermitian physics.

Genuinely non-Hermitian dynamical signatures require the quench to change the metric, i.e., $\eta_i\neq\eta_f$.  In that case the physical inner product itself changes at $t=0$, the doubled-space inner product would involve both $\eta_i$ and $\eta_f$, and the Loschmidt amplitude would no longer be given by a simple trace formula.  Interesting phenomena could then emerge.  Such metric-changing quenches lie beyond the scope of the present work and constitute an important direction for future research.

\section{Conclusion}\label{Sec5}

In this work we have established a self-consistent framework for finite-temperature dynamical quantum phase transitions in quasi-Hermitian systems.  Starting from the basic tenets of pseudo-Hermitian quantum mechanics, we derived the mixed-state Loschmidt amplitude from first principles using purification under a metric-stationary quench.  Applying this framework to an $N$-level model with a dynamical two-level sector and spectator states, we identified a universal thermal-suppression mechanism governed by the redistribution of statistical weights among the eigenstates.  A dimensionality-dependent threshold temperature emerges, which becomes finite for $N\ge 5$, above which DQPTs are fully suppressed.  This transition is purely dynamical, with no underlying equilibrium phase transition.

A central result of this work is the strict equivalence between the quasi-Hermitian Loschmidt amplitude and that of the associated Hermitian system under a fixed positive-definite metric.  Within this metric-stationary framework, non-Hermiticity is absorbed into a global similarity transformation and manifests only as a renormalization of energy scales.  While the thermal suppression mechanism is therefore universal, the quasi-Hermitian framework was indispensable for its derivation, as it supplies the consistent probabilistic interpretation and well-defined purification structure needed to construct the Loschmidt amplitude in systems with non-orthogonal eigenstates.  Without it, the very definition of the mixed-state Loschmidt echo in a biorthogonal system would remain ambiguous.  Genuinely non-Hermitian dynamical signatures, such as complex interference terms and parameter-space holonomies, require relaxing the fixed-metric assumption; such metric-changing protocols constitute an important direction for future work.

\begin{acknowledgments}
H. G. was supported by the Quantum Science and Technology-National Science and Technology Major Project (Grant No. 2021ZD0301904) and the National Natural Science Foundation of China (Grant No. 12447216). X. Y. H. was supported by the Jiangsu Funding Program for Excellent Postdoctoral Talent (Grant No. 2023ZB611).

Jia-Chen Tang and Xu-Yang Hou contributed equally
to this work.
\end{acknowledgments}

\appendix
\section{Proof of the purification inner-product identity}
\label{purification-inner}

We show that the $\eta$-Hilbert-Schmidt inner product of two purifications equals the physical inner product of the corresponding purified states in the doubled space.  Writing $W_j = \sum_n \sqrt{\lambda_n^{(j)}} |\psi_n\rangle \langle \phi_n|\, \mathcal{U}_j$ and using the biorthogonality relations, we have
\begin{align}
\mathrm{Tr}(W_1^{\ddag} W_2)
 &= \mathrm{Tr}\!\Big( \sum_n \mathcal{U}_1^{\ddag} \sqrt{\lambda_n^{(1)}}
      \eta^{-1} |\phi_n\rangle \langle \psi_n| \eta \notag\\
      &\qquad\times\sum_m \sqrt{\lambda_m^{(2)}} |\psi_m\rangle \langle \phi_m| \mathcal{U}_2 \Big) \notag \\
 &= \mathrm{Tr}\!\left( \sum_{n,m} \sqrt{\lambda_n^{(1)}\lambda_m^{(2)}} \,
      \mathcal{U}_1^{\ddag} |\psi_n\rangle \langle \phi_n|\psi_m\rangle \langle \phi_m|\, \mathcal{U}_2 \right) \notag \\
% &= \mathrm{Tr}\!\left( \sum_n \sqrt{\lambda_n^{(1)}\lambda_n^{(2)}} \,
%      \mathcal{U}_1^{\ddag} |\psi_n\rangle \langle \phi_n|\, \mathcal{U}_2 \right) \notag \\
 &= \mathrm{Tr}\!\left( \sum_n \sqrt{\lambda_n^{(1)}\lambda_n^{(2)}} \,
      |\psi_n\rangle \langle \phi_n|\, \mathcal{U}_2 \mathcal{U}_1^{\ddag} \right). \notag
\end{align}
On the other hand, the doubled-space inner product gives
\begin{align}
\langle W_1 | W_2 \rangle_{\eta}
 &= \sum_{n,m} \sqrt{\lambda_n^{(1)}\lambda_m^{(2)}} \,
    \langle \psi_n | \eta | \psi_m \rangle \,
    \langle \phi_m | \mathcal{U}_2 \,\eta^{-1} \mathcal{U}_1^\dagger | \phi_n \rangle \notag \\
% &= \sum_{n,m} \sqrt{\lambda_n^{(1)}\lambda_m^{(2)}} \, \delta_{mn} \,
%    \langle \phi_m | \mathcal{U}_2 \,\eta^{-1} \mathcal{U}_1^\dagger \eta | \psi_n \rangle \notag \\
 &= \sum_n \sqrt{\lambda_n^{(1)}\lambda_n^{(2)}} \,
    \langle \phi_n | \mathcal{U}_2 \, \mathcal{U}_1^{\ddag} | \psi_n \rangle \notag \\
 &= \mathrm{Tr}\!\left( \sum_n \sqrt{\lambda_n^{(1)}\lambda_n^{(2)}} \,
      |\psi_n\rangle \langle \phi_n|\, \mathcal{U}_2 \mathcal{U}_1^{\ddag} \right),
\end{align}
which coincides with the trace expression above.  Hence $\mathrm{Tr}(W_1^{\ddag} W_2) = \langle W_1 | W_2 \rangle_{\eta}$.

\section{No obstructions to Hermitianization in the metric-stationary quench}
\label{app2}

We show that the quench protocol studied in the main text is free from both geometric and topological obstructions to Hermitianization.  The analysis is carried out for the two-level sector $H_2=\boldsymbol{R}\cdot\boldsymbol{\sigma}$ with $\boldsymbol{R}=a\,\boldsymbol{n}^r+\mathrm{i}b\,\boldsymbol{n}^\theta$.

A positive-definite metric operator satisfying the quasi-Hermiticity condition $H_2^\dagger\eta=\eta H_2$ can be chosen as $\eta = 1_2 - \frac{b}{a}\,\sigma_\phi$,
where $\sigma_\phi=-\sin\phi\,\sigma_x+\cos\phi\,\sigma_y$.  The eigenvalues of $\eta$ are $1\pm b/a$, which are strictly positive in the unbroken $\mathcal{PT}$ regime $a^2>b^2$.  The metric adopted in the main text is merely a positive constant multiple of this operator; since the connection $\mathcal{G}_\mu=\frac{1}{2}[\partial_\mu\sqrt{\eta},\sqrt{\eta}^{-1}]$ is invariant under an overall rescaling of $\eta$, both choices yield identical geometric data.
Because $\eta$ depends only on the azimuthal angle $\phi$ (through $\sigma_\phi$) and on the ratio $b/a$, but not on the polar angle $\theta$, we have $\partial_\theta\eta=0$ and consequently $\partial_\theta\sqrt{\eta}=0$.  The gauge connection therefore has a vanishing $\theta$ component $\mathcal{G}_\theta = \frac{1}{2}\bigl[\partial_\theta\sqrt{\eta},\sqrt{\eta}^{-1}\bigr] = 0$.
For the $\phi$ component, writing $\sqrt{\eta}=c_01_2+c_1\sigma_\phi$ with $c_0 = \sqrt{\frac{1+\sqrt{1-b^2/a^2}}{2}}$, $c_1 = -\frac{b}{2a c_0}$ 
and using $\partial_\phi\sigma_\phi=-\sigma_\rho$ with $\sigma_\rho=\cos\phi\,\sigma_x+\sin\phi\,\sigma_y$, one finds
\begin{align}
\mathcal{G}_\phi = \frac{1}{2}\bigl[\partial_\phi\sqrt{\eta},\sqrt{\eta}^{-1}\bigr]
= \mathrm{i}K\sigma_z,
\end{align}
where $K=\frac{1}{2}\left(\frac{a}{\sqrt{a^2-b^2}}-1\right)$.
The curvature of the connection is
\begin{align}
\mathcal{F}_{\theta\phi}^{\mathcal{G}}
= \partial_\theta\mathcal{G}_\phi - \partial_\phi\mathcal{G}_\theta - [\mathcal{G}_\theta,\mathcal{G}_\phi]
= 0,
\end{align}
because $\mathcal{G}_\theta=0$ and $\mathcal{G}_\phi$ is independent of $\theta$.  Thus the geometric obstruction vanishes identically.

In the quench protocol, the metric-stationary condition requires $\phi_i=\phi_f$ and fixes the ratio $b/a$, so the system does not traverse the $\phi$ direction.  Since the only non-vanishing connection component is $\mathcal{G}_\phi$ and the quench path lies entirely at fixed $\phi$, the parallel-transport equation $\partial_\mu U = -U\mathcal{G}_\mu$ generates only the trivial solution along the physical trajectory.  Furthermore, because the quench is an open path (not a closed loop) in parameter space, no non-contractible cycle is available to support a nontrivial Wilson loop.  Hence both geometric and topological obstructions are absent, and the proper similarity transformation $S_p$ can be chosen globally and single-valued for the quench dynamics.


\begin{thebibliography}{61}
\expandafter\ifx\csname natexlab\endcsname\relax\def\natexlab#1{#1}\fi
\expandafter\ifx\csname bibnamefont\endcsname\relax
  \def\bibnamefont#1{#1}\fi
\expandafter\ifx\csname bibfnamefont\endcsname\relax
  \def\bibfnamefont#1{#1}\fi
\expandafter\ifx\csname citenamefont\endcsname\relax
  \def\citenamefont#1{#1}\fi
\expandafter\ifx\csname url\endcsname\relax
  \def\url#1{\texttt{#1}}\fi
\expandafter\ifx\csname urlprefix\endcsname\relax\def\urlprefix{URL }\fi
\providecommand{\bibinfo}[2]{#2}
\providecommand{\eprint}[2][]{\url{#2}}

\bibitem[{\citenamefont{Heyl et~al.}(2013)\citenamefont{Heyl, Polkovnikov, and
  Kehrein}}]{Heyl2013}
\bibinfo{author}{\bibfnamefont{M.}~\bibnamefont{Heyl}},
  \bibinfo{author}{\bibfnamefont{A.}~\bibnamefont{Polkovnikov}},
  \bibnamefont{and} \bibinfo{author}{\bibfnamefont{S.}~\bibnamefont{Kehrein}},
  \bibinfo{journal}{Phys. Rev. Lett.} \textbf{\bibinfo{volume}{110}},
  \bibinfo{pages}{135704} (\bibinfo{year}{2013}).

\bibitem[{\citenamefont{Heyl}(2018)}]{Heyl2018}
\bibinfo{author}{\bibfnamefont{M.}~\bibnamefont{Heyl}}, \bibinfo{journal}{Rep.
  Prog. Phys.} \textbf{\bibinfo{volume}{81}}, \bibinfo{pages}{054001}
  (\bibinfo{year}{2018}).

\bibitem[{\citenamefont{Heyl}(2014)}]{Heyl2014}
\bibinfo{author}{\bibfnamefont{M.}~\bibnamefont{Heyl}}, \bibinfo{journal}{Phys.
  Rev. Lett.} \textbf{\bibinfo{volume}{113}}, \bibinfo{pages}{205701}
  (\bibinfo{year}{2014}).

\bibitem[{\citenamefont{Heyl}(2015)}]{Heyl2015}
\bibinfo{author}{\bibfnamefont{M.}~\bibnamefont{Heyl}}, \bibinfo{journal}{Phys.
  Rev. Lett.} \textbf{\bibinfo{volume}{115}}, \bibinfo{pages}{140602}
  (\bibinfo{year}{2015}).

\bibitem[{\citenamefont{Budich and Heyl}(2016)}]{Budich2016}
\bibinfo{author}{\bibfnamefont{J.~C.} \bibnamefont{Budich}} \bibnamefont{and}
  \bibinfo{author}{\bibfnamefont{M.}~\bibnamefont{Heyl}},
  \bibinfo{journal}{Phys. Rev. B} \textbf{\bibinfo{volume}{93}},
  \bibinfo{pages}{085416} (\bibinfo{year}{2016}).

\bibitem[{\citenamefont{Vajna and Dora}(2014)}]{Vajna2015a}
\bibinfo{author}{\bibfnamefont{S.}~\bibnamefont{Vajna}} \bibnamefont{and}
  \bibinfo{author}{\bibfnamefont{B.}~\bibnamefont{Dora}},
  \bibinfo{journal}{Phys. Rev. B} \textbf{\bibinfo{volume}{89}},
  \bibinfo{pages}{161105(R)} (\bibinfo{year}{2014}).

\bibitem[{\citenamefont{Vajna and Dora}(2015)}]{Vajna2015b}
\bibinfo{author}{\bibfnamefont{S.}~\bibnamefont{Vajna}} \bibnamefont{and}
  \bibinfo{author}{\bibfnamefont{B.}~\bibnamefont{Dora}},
  \bibinfo{journal}{Phys. Rev. B} \textbf{\bibinfo{volume}{91}},
  \bibinfo{pages}{155127} (\bibinfo{year}{2015}).

\bibitem[{\citenamefont{Bhattacharya et~al.}(2017)\citenamefont{Bhattacharya,
  Bandyopadhyay, and Dutta}}]{Bhattacharya2017}
\bibinfo{author}{\bibfnamefont{U.}~\bibnamefont{Bhattacharya}},
  \bibinfo{author}{\bibfnamefont{S.}~\bibnamefont{Bandyopadhyay}},
  \bibnamefont{and} \bibinfo{author}{\bibfnamefont{A.}~\bibnamefont{Dutta}},
  \bibinfo{journal}{Phys. Rev. B} \textbf{\bibinfo{volume}{96}},
  \bibinfo{pages}{180303(R)} (\bibinfo{year}{2017}).

\bibitem[{\citenamefont{Heyl and Budich}(2017)}]{Heyl2017}
\bibinfo{author}{\bibfnamefont{M.}~\bibnamefont{Heyl}} \bibnamefont{and}
  \bibinfo{author}{\bibfnamefont{J.~C.} \bibnamefont{Budich}},
  \bibinfo{journal}{Phys. Rev. B} \textbf{\bibinfo{volume}{96}},
  \bibinfo{pages}{180304(R)} (\bibinfo{year}{2017}).

\bibitem[{\citenamefont{Hou et~al.}(2020)\citenamefont{Hou, Gao, Guo, He, Liu,
  and Chien}}]{Hou2020a}
\bibinfo{author}{\bibfnamefont{X.-Y.} \bibnamefont{Hou}},
  \bibinfo{author}{\bibfnamefont{Q.-C.} \bibnamefont{Gao}},
  \bibinfo{author}{\bibfnamefont{H.}~\bibnamefont{Guo}},
  \bibinfo{author}{\bibfnamefont{Y.}~\bibnamefont{He}},
  \bibinfo{author}{\bibfnamefont{T.}~\bibnamefont{Liu}}, \bibnamefont{and}
  \bibinfo{author}{\bibfnamefont{C.-C.} \bibnamefont{Chien}},
  \bibinfo{journal}{Phys. Rev. B} \textbf{\bibinfo{volume}{102}},
  \bibinfo{pages}{104305} (\bibinfo{year}{2020}).

\bibitem[{\citenamefont{Hou et~al.}(2022)\citenamefont{Hou, Gao, Guo, and
  Chien}}]{Hou2022}
\bibinfo{author}{\bibfnamefont{X.-Y.} \bibnamefont{Hou}},
  \bibinfo{author}{\bibfnamefont{Q.-C.} \bibnamefont{Gao}},
  \bibinfo{author}{\bibfnamefont{H.}~\bibnamefont{Guo}}, \bibnamefont{and}
  \bibinfo{author}{\bibfnamefont{C.-C.} \bibnamefont{Chien}},
  \bibinfo{journal}{Phys. Rev. B} \textbf{\bibinfo{volume}{106}},
  \bibinfo{pages}{014301} (\bibinfo{year}{2022}).

\bibitem[{\citenamefont{Sedlmayr et~al.}(2018)\citenamefont{Sedlmayr,
  Fleischhauer, and Sirker}}]{Sedlmayr2018}
\bibinfo{author}{\bibfnamefont{N.}~\bibnamefont{Sedlmayr}},
  \bibinfo{author}{\bibfnamefont{M.}~\bibnamefont{Fleischhauer}},
  \bibnamefont{and} \bibinfo{author}{\bibfnamefont{J.}~\bibnamefont{Sirker}},
  \bibinfo{journal}{Phys. Rev. B} \textbf{\bibinfo{volume}{97}},
  \bibinfo{pages}{045147} (\bibinfo{year}{2018}).

\bibitem[{\citenamefont{Mera et~al.}(2018)\citenamefont{Mera, Vlachou,
  Paunkovi{\'c}, Vieira, and Viyuela}}]{Mera2018}
\bibinfo{author}{\bibfnamefont{B.}~\bibnamefont{Mera}},
  \bibinfo{author}{\bibfnamefont{C.}~\bibnamefont{Vlachou}},
  \bibinfo{author}{\bibfnamefont{N.}~\bibnamefont{Paunkovi{\'c}}},
  \bibinfo{author}{\bibfnamefont{V.~R.} \bibnamefont{Vieira}},
  \bibnamefont{and} \bibinfo{author}{\bibfnamefont{O.}~\bibnamefont{Viyuela}},
  \bibinfo{journal}{Phys. Rev. B} \textbf{\bibinfo{volume}{97}},
  \bibinfo{pages}{094110} (\bibinfo{year}{2018}).

\bibitem[{\citenamefont{Lang et~al.}(2018{\natexlab{a}})\citenamefont{Lang,
  Frank, and Halimeh}}]{Lang2018a}
\bibinfo{author}{\bibfnamefont{J.}~\bibnamefont{Lang}},
  \bibinfo{author}{\bibfnamefont{B.}~\bibnamefont{Frank}}, \bibnamefont{and}
  \bibinfo{author}{\bibfnamefont{J.~C.} \bibnamefont{Halimeh}},
  \bibinfo{journal}{Phys. Rev. Lett.} \textbf{\bibinfo{volume}{121}},
  \bibinfo{pages}{130603} (\bibinfo{year}{2018}{\natexlab{a}}).

\bibitem[{\citenamefont{Lang et~al.}(2018{\natexlab{b}})\citenamefont{Lang,
  Frank, and Halimeh}}]{Lang2018b}
\bibinfo{author}{\bibfnamefont{J.}~\bibnamefont{Lang}},
  \bibinfo{author}{\bibfnamefont{B.}~\bibnamefont{Frank}}, \bibnamefont{and}
  \bibinfo{author}{\bibfnamefont{J.~C.} \bibnamefont{Halimeh}},
  \bibinfo{journal}{Phys. Rev. B} \textbf{\bibinfo{volume}{97}},
  \bibinfo{pages}{174401} (\bibinfo{year}{2018}{\natexlab{b}}).

\bibitem[{\citenamefont{Halimeh and Zauner-Stauber}(2017)}]{Halimeh2017}
\bibinfo{author}{\bibfnamefont{J.~C.} \bibnamefont{Halimeh}} \bibnamefont{and}
  \bibinfo{author}{\bibfnamefont{V.}~\bibnamefont{Zauner-Stauber}},
  \bibinfo{journal}{Phys. Rev. B} \textbf{\bibinfo{volume}{96}},
  \bibinfo{pages}{134427} (\bibinfo{year}{2017}).

\bibitem[{\citenamefont{Homrighausen et~al.}(2017)\citenamefont{Homrighausen,
  Abeling, Zauner-Stauber, and Halimeh}}]{Homrighausen2017}
\bibinfo{author}{\bibfnamefont{I.}~\bibnamefont{Homrighausen}},
  \bibinfo{author}{\bibfnamefont{N.~O.} \bibnamefont{Abeling}},
  \bibinfo{author}{\bibfnamefont{V.}~\bibnamefont{Zauner-Stauber}},
  \bibnamefont{and} \bibinfo{author}{\bibfnamefont{J.~C.}
  \bibnamefont{Halimeh}}, \bibinfo{journal}{Phys. Rev. B}
  \textbf{\bibinfo{volume}{96}}, \bibinfo{pages}{104436}
  (\bibinfo{year}{2017}).

\bibitem[{\citenamefont{Zunkovic et~al.}(2018)\citenamefont{Zunkovic, Heyl,
  Knap, and Silva}}]{Zunkovic2018}
\bibinfo{author}{\bibfnamefont{B.}~\bibnamefont{Zunkovic}},
  \bibinfo{author}{\bibfnamefont{M.}~\bibnamefont{Heyl}},
  \bibinfo{author}{\bibfnamefont{M.}~\bibnamefont{Knap}}, \bibnamefont{and}
  \bibinfo{author}{\bibfnamefont{A.}~\bibnamefont{Silva}},
  \bibinfo{journal}{Phys. Rev. Lett.} \textbf{\bibinfo{volume}{120}},
  \bibinfo{pages}{130601} (\bibinfo{year}{2018}).

\bibitem[{\citenamefont{Karrasch and Schuricht}(2013)}]{Karrasch2013}
\bibinfo{author}{\bibfnamefont{C.}~\bibnamefont{Karrasch}} \bibnamefont{and}
  \bibinfo{author}{\bibfnamefont{D.}~\bibnamefont{Schuricht}},
  \bibinfo{journal}{Phys. Rev. B} \textbf{\bibinfo{volume}{87}},
  \bibinfo{pages}{195104} (\bibinfo{year}{2013}).

\bibitem[{\citenamefont{Kriel et~al.}(2014)\citenamefont{Kriel, Karrasch, and
  Kehrein}}]{Karrasch2014}
\bibinfo{author}{\bibfnamefont{J.~N.} \bibnamefont{Kriel}},
  \bibinfo{author}{\bibfnamefont{C.}~\bibnamefont{Karrasch}}, \bibnamefont{and}
  \bibinfo{author}{\bibfnamefont{S.}~\bibnamefont{Kehrein}},
  \bibinfo{journal}{Phys. Rev. B} \textbf{\bibinfo{volume}{90}},
  \bibinfo{pages}{125106} (\bibinfo{year}{2014}).

\bibitem[{\citenamefont{Andraschko and Sirker}(2014)}]{Andraschko2014}
\bibinfo{author}{\bibfnamefont{F.}~\bibnamefont{Andraschko}} \bibnamefont{and}
  \bibinfo{author}{\bibfnamefont{J.}~\bibnamefont{Sirker}},
  \bibinfo{journal}{Phys. Rev. B} \textbf{\bibinfo{volume}{89}},
  \bibinfo{pages}{125120} (\bibinfo{year}{2014}).

\bibitem[{\citenamefont{Schmitt and Kehrein}(2015)}]{Schmitt2015}
\bibinfo{author}{\bibfnamefont{M.}~\bibnamefont{Schmitt}} \bibnamefont{and}
  \bibinfo{author}{\bibfnamefont{S.}~\bibnamefont{Kehrein}},
  \bibinfo{journal}{Phys. Rev. B} \textbf{\bibinfo{volume}{92}},
  \bibinfo{pages}{075114} (\bibinfo{year}{2015}).

\bibitem[{\citenamefont{Sharma et~al.}(2015)\citenamefont{Sharma, Suzuki, and
  Dutta}}]{Sharma2015}
\bibinfo{author}{\bibfnamefont{S.}~\bibnamefont{Sharma}},
  \bibinfo{author}{\bibfnamefont{S.}~\bibnamefont{Suzuki}}, \bibnamefont{and}
  \bibinfo{author}{\bibfnamefont{A.}~\bibnamefont{Dutta}},
  \bibinfo{journal}{Phys. Rev. B} \textbf{\bibinfo{volume}{92}},
  \bibinfo{pages}{104306} (\bibinfo{year}{2015}).

\bibitem[{\citenamefont{Divakaran et~al.}(2016)\citenamefont{Divakaran, Sharma,
  and Dutta}}]{Divakaran2016}
\bibinfo{author}{\bibfnamefont{U.}~\bibnamefont{Divakaran}},
  \bibinfo{author}{\bibfnamefont{S.}~\bibnamefont{Sharma}}, \bibnamefont{and}
  \bibinfo{author}{\bibfnamefont{A.}~\bibnamefont{Dutta}},
  \bibinfo{journal}{Phys. Rev. E} \textbf{\bibinfo{volume}{93}},
  \bibinfo{pages}{052133} (\bibinfo{year}{2016}).

\bibitem[{\citenamefont{Zamani et~al.}(2020)\citenamefont{Zamani, Jafari, and
  Langari}}]{Zamani2020}
\bibinfo{author}{\bibfnamefont{S.}~\bibnamefont{Zamani}},
  \bibinfo{author}{\bibfnamefont{R.}~\bibnamefont{Jafari}}, \bibnamefont{and}
  \bibinfo{author}{\bibfnamefont{A.}~\bibnamefont{Langari}},
  \bibinfo{journal}{Phys. Rev. B} \textbf{\bibinfo{volume}{102}},
  \bibinfo{pages}{144306} (\bibinfo{year}{2020}).

\bibitem[{\citenamefont{Jafari and Akbari}(2021)}]{Jafari2021a}
\bibinfo{author}{\bibfnamefont{R.}~\bibnamefont{Jafari}} \bibnamefont{and}
  \bibinfo{author}{\bibfnamefont{A.}~\bibnamefont{Akbari}},
  \bibinfo{journal}{Phys. Rev. A} \textbf{\bibinfo{volume}{103}},
  \bibinfo{pages}{012204} (\bibinfo{year}{2021}).

\bibitem[{\citenamefont{Jafari et~al.}(2022)\citenamefont{Jafari, Akbari,
  Mishra, and Johannesson}}]{Jafari2021b}
\bibinfo{author}{\bibfnamefont{R.}~\bibnamefont{Jafari}},
  \bibinfo{author}{\bibfnamefont{A.}~\bibnamefont{Akbari}},
  \bibinfo{author}{\bibfnamefont{U.}~\bibnamefont{Mishra}}, \bibnamefont{and}
  \bibinfo{author}{\bibfnamefont{H.}~\bibnamefont{Johannesson}},
  \bibinfo{journal}{Phys. Rev. B} \textbf{\bibinfo{volume}{105}},
  \bibinfo{pages}{094311} (\bibinfo{year}{2022}).

\bibitem[{\citenamefont{Zamani et~al.}(2022)\citenamefont{Zamani, Jafari, and
  Langari}}]{Jafari2022a}
\bibinfo{author}{\bibfnamefont{S.}~\bibnamefont{Zamani}},
  \bibinfo{author}{\bibfnamefont{R.}~\bibnamefont{Jafari}}, \bibnamefont{and}
  \bibinfo{author}{\bibfnamefont{A.}~\bibnamefont{Langari}},
  \bibinfo{journal}{Phys. Rev. B} \textbf{\bibinfo{volume}{105}},
  \bibinfo{pages}{094304} (\bibinfo{year}{2022}).

\bibitem[{\citenamefont{Naji et~al.}(2022)\citenamefont{Naji, Jafari, Jafari,
  and Akbari}}]{Naji2022}
\bibinfo{author}{\bibfnamefont{J.}~\bibnamefont{Naji}},
  \bibinfo{author}{\bibfnamefont{M.}~\bibnamefont{Jafari}},
  \bibinfo{author}{\bibfnamefont{R.}~\bibnamefont{Jafari}}, \bibnamefont{and}
  \bibinfo{author}{\bibfnamefont{A.}~\bibnamefont{Akbari}},
  \bibinfo{journal}{Phys. Rev. A} \textbf{\bibinfo{volume}{105}},
  \bibinfo{pages}{022220} (\bibinfo{year}{2022}).

\bibitem[{\citenamefont{Yang et~al.}(2019)\citenamefont{Yang, Zhou, Ma, Kong,
  Wang, Qin, Rong, Wang, Shi, Gong et~al.}}]{Yang2019}
\bibinfo{author}{\bibfnamefont{K.}~\bibnamefont{Yang}},
  \bibinfo{author}{\bibfnamefont{L.}~\bibnamefont{Zhou}},
  \bibinfo{author}{\bibfnamefont{W.}~\bibnamefont{Ma}},
  \bibinfo{author}{\bibfnamefont{X.}~\bibnamefont{Kong}},
  \bibinfo{author}{\bibfnamefont{P.}~\bibnamefont{Wang}},
  \bibinfo{author}{\bibfnamefont{X.}~\bibnamefont{Qin}},
  \bibinfo{author}{\bibfnamefont{X.}~\bibnamefont{Rong}},
  \bibinfo{author}{\bibfnamefont{Y.}~\bibnamefont{Wang}},
  \bibinfo{author}{\bibfnamefont{F.}~\bibnamefont{Shi}},
  \bibinfo{author}{\bibfnamefont{J.}~\bibnamefont{Gong}}, \bibnamefont{et~al.},
  \bibinfo{journal}{Phys. Rev. B} \textbf{\bibinfo{volume}{100}},
  \bibinfo{pages}{085308} (\bibinfo{year}{2019}).

\bibitem[{\citenamefont{Zhou and Du}(2021{\natexlab{a}})}]{Zhou2021a}
\bibinfo{author}{\bibfnamefont{L.}~\bibnamefont{Zhou}} \bibnamefont{and}
  \bibinfo{author}{\bibfnamefont{Q.}~\bibnamefont{Du}}, \bibinfo{journal}{J.
  Phys.: Condens. Matter} \textbf{\bibinfo{volume}{33}},
  \bibinfo{pages}{345403} (\bibinfo{year}{2021}{\natexlab{a}}).

\bibitem[{\citenamefont{Hamazaki}(2021)}]{Hamazaki2021}
\bibinfo{author}{\bibfnamefont{R.}~\bibnamefont{Hamazaki}},
  \bibinfo{journal}{Nat. Commun.} \textbf{\bibinfo{volume}{12}},
  \bibinfo{pages}{5108} (\bibinfo{year}{2021}).

\bibitem[{\citenamefont{Ashida et~al.}(2020)\citenamefont{Ashida, Gong, and
  Ueda}}]{Ashida2020}
\bibinfo{author}{\bibfnamefont{Y.}~\bibnamefont{Ashida}},
  \bibinfo{author}{\bibfnamefont{Z.}~\bibnamefont{Gong}}, \bibnamefont{and}
  \bibinfo{author}{\bibfnamefont{M.}~\bibnamefont{Ueda}},
  \bibinfo{journal}{Adv. Phys.} \textbf{\bibinfo{volume}{69}},
  \bibinfo{pages}{249} (\bibinfo{year}{2020}).

\bibitem[{\citenamefont{El-Ganainy et~al.}(2018)\citenamefont{El-Ganainy,
  Makris, Khajavikhan, Musslimani, Rotter, and
  Christodoulides}}]{El-Ganainy2018}
\bibinfo{author}{\bibfnamefont{R.}~\bibnamefont{El-Ganainy}},
  \bibinfo{author}{\bibfnamefont{K.~G.} \bibnamefont{Makris}},
  \bibinfo{author}{\bibfnamefont{M.}~\bibnamefont{Khajavikhan}},
  \bibinfo{author}{\bibfnamefont{Z.~H.} \bibnamefont{Musslimani}},
  \bibinfo{author}{\bibfnamefont{S.}~\bibnamefont{Rotter}}, \bibnamefont{and}
  \bibinfo{author}{\bibfnamefont{D.~N.} \bibnamefont{Christodoulides}},
  \bibinfo{journal}{Nat. Phys.} \textbf{\bibinfo{volume}{14}},
  \bibinfo{pages}{11} (\bibinfo{year}{2018}).

\bibitem[{\citenamefont{Bender and Boettcher}(1998)}]{Bender1998}
\bibinfo{author}{\bibfnamefont{C.~M.} \bibnamefont{Bender}} \bibnamefont{and}
  \bibinfo{author}{\bibfnamefont{S.}~\bibnamefont{Boettcher}},
  \bibinfo{journal}{Phys. Rev. Lett.} \textbf{\bibinfo{volume}{80}},
  \bibinfo{pages}{5243} (\bibinfo{year}{1998}).

\bibitem[{\citenamefont{Bergholtz et~al.}(2021)\citenamefont{Bergholtz, Budich,
  and Kunst}}]{Bergholtz2021}
\bibinfo{author}{\bibfnamefont{E.~J.} \bibnamefont{Bergholtz}},
  \bibinfo{author}{\bibfnamefont{J.~C.} \bibnamefont{Budich}},
  \bibnamefont{and} \bibinfo{author}{\bibfnamefont{F.~K.} \bibnamefont{Kunst}},
  \bibinfo{journal}{Rev. Mod. Phys.} \textbf{\bibinfo{volume}{93}},
  \bibinfo{pages}{015005} (\bibinfo{year}{2021}).

\bibitem[{\citenamefont{Gong et~al.}(2018)\citenamefont{Gong, Ashida, Kawabata,
  Takasan, Higashikawa, and Ueda}}]{Gong2018}
\bibinfo{author}{\bibfnamefont{Z.}~\bibnamefont{Gong}},
  \bibinfo{author}{\bibfnamefont{Y.}~\bibnamefont{Ashida}},
  \bibinfo{author}{\bibfnamefont{K.}~\bibnamefont{Kawabata}},
  \bibinfo{author}{\bibfnamefont{K.}~\bibnamefont{Takasan}},
  \bibinfo{author}{\bibfnamefont{S.}~\bibnamefont{Higashikawa}},
  \bibnamefont{and} \bibinfo{author}{\bibfnamefont{M.}~\bibnamefont{Ueda}},
  \bibinfo{journal}{Phys. Rev. X} \textbf{\bibinfo{volume}{8}},
  \bibinfo{pages}{031079} (\bibinfo{year}{2018}).

\bibitem[{\citenamefont{Kunst et~al.}(2018)\citenamefont{Kunst, Edvardsson,
  Budich, and Bergholtz}}]{Kunst2018}
\bibinfo{author}{\bibfnamefont{F.~K.} \bibnamefont{Kunst}},
  \bibinfo{author}{\bibfnamefont{E.}~\bibnamefont{Edvardsson}},
  \bibinfo{author}{\bibfnamefont{J.~C.} \bibnamefont{Budich}},
  \bibnamefont{and} \bibinfo{author}{\bibfnamefont{E.~J.}
  \bibnamefont{Bergholtz}}, \bibinfo{journal}{Phys. Rev. Lett.}
  \textbf{\bibinfo{volume}{121}}, \bibinfo{pages}{026808}
  (\bibinfo{year}{2018}).

\bibitem[{\citenamefont{Yao and Wang}(2018)}]{Yao2018}
\bibinfo{author}{\bibfnamefont{S.}~\bibnamefont{Yao}} \bibnamefont{and}
  \bibinfo{author}{\bibfnamefont{Z.}~\bibnamefont{Wang}},
  \bibinfo{journal}{Phys. Rev. Lett.} \textbf{\bibinfo{volume}{121}},
  \bibinfo{pages}{086803} (\bibinfo{year}{2018}).

\bibitem[{\citenamefont{Kawabata et~al.}(2019)\citenamefont{Kawabata, Shiozaki,
  Ueda, and Sato}}]{Kawabata2019}
\bibinfo{author}{\bibfnamefont{K.}~\bibnamefont{Kawabata}},
  \bibinfo{author}{\bibfnamefont{K.}~\bibnamefont{Shiozaki}},
  \bibinfo{author}{\bibfnamefont{M.}~\bibnamefont{Ueda}}, \bibnamefont{and}
  \bibinfo{author}{\bibfnamefont{M.}~\bibnamefont{Sato}},
  \bibinfo{journal}{Phys. Rev. X} \textbf{\bibinfo{volume}{9}},
  \bibinfo{pages}{041015} (\bibinfo{year}{2019}).

\bibitem[{\citenamefont{Brody}(2014)}]{Brody2014}
\bibinfo{author}{\bibfnamefont{D.~C.} \bibnamefont{Brody}},
  \bibinfo{journal}{J. Phys. A: Math. Theor.} \textbf{\bibinfo{volume}{47}},
  \bibinfo{pages}{035305} (\bibinfo{year}{2014}).

\bibitem[{\citenamefont{Tang et~al.}(2022)\citenamefont{Tang, Kou, and
  Sun}}]{Tang2022Biorthogonal}
\bibinfo{author}{\bibfnamefont{J.-C.} \bibnamefont{Tang}},
  \bibinfo{author}{\bibfnamefont{S.-P.} \bibnamefont{Kou}}, \bibnamefont{and}
  \bibinfo{author}{\bibfnamefont{G.}~\bibnamefont{Sun}},
  \bibinfo{journal}{Europhys. Lett.} \textbf{\bibinfo{volume}{137}},
  \bibinfo{pages}{40001} (\bibinfo{year}{2022}).

\bibitem[{\citenamefont{Sun et~al.}(2022)\citenamefont{Sun, Tang, and
  Kou}}]{Sun2022Biorthogonal}
\bibinfo{author}{\bibfnamefont{G.}~\bibnamefont{Sun}},
  \bibinfo{author}{\bibfnamefont{J.-C.} \bibnamefont{Tang}}, \bibnamefont{and}
  \bibinfo{author}{\bibfnamefont{S.-P.} \bibnamefont{Kou}},
  \bibinfo{journal}{Front. Phys.} \textbf{\bibinfo{volume}{17}},
  \bibinfo{pages}{33502} (\bibinfo{year}{2022}).

\bibitem[{\citenamefont{Mondal and Nag}(2022)}]{Mondal2022}
\bibinfo{author}{\bibfnamefont{D.}~\bibnamefont{Mondal}} \bibnamefont{and}
  \bibinfo{author}{\bibfnamefont{T.}~\bibnamefont{Nag}},
  \bibinfo{journal}{Phys. Rev. B} \textbf{\bibinfo{volume}{106}},
  \bibinfo{pages}{054308} (\bibinfo{year}{2022}).

\bibitem[{\citenamefont{Mondal and Nag}(2023)}]{Mondal2023PhysRevB}
\bibinfo{author}{\bibfnamefont{D.}~\bibnamefont{Mondal}} \bibnamefont{and}
  \bibinfo{author}{\bibfnamefont{T.}~\bibnamefont{Nag}},
  \bibinfo{journal}{Phys. Rev. B} \textbf{\bibinfo{volume}{107}},
  \bibinfo{pages}{184311} (\bibinfo{year}{2023}).

\bibitem[{\citenamefont{Zhou et~al.}(2018)\citenamefont{Zhou, Wang, Wang, and
  Gong}}]{Zhou2018}
\bibinfo{author}{\bibfnamefont{L.}~\bibnamefont{Zhou}},
  \bibinfo{author}{\bibfnamefont{Q.-h.} \bibnamefont{Wang}},
  \bibinfo{author}{\bibfnamefont{H.}~\bibnamefont{Wang}}, \bibnamefont{and}
  \bibinfo{author}{\bibfnamefont{J.}~\bibnamefont{Gong}},
  \bibinfo{journal}{Phys. Rev. A} \textbf{\bibinfo{volume}{98}},
  \bibinfo{pages}{022129} (\bibinfo{year}{2018}).

\bibitem[{\citenamefont{Zhou and Du}(2021{\natexlab{b}})}]{Zhou2021b}
\bibinfo{author}{\bibfnamefont{L.}~\bibnamefont{Zhou}} \bibnamefont{and}
  \bibinfo{author}{\bibfnamefont{Q.}~\bibnamefont{Du}}, \bibinfo{journal}{New
  J. Phys.} \textbf{\bibinfo{volume}{23}}, \bibinfo{pages}{063041}
  (\bibinfo{year}{2021}{\natexlab{b}}).

\bibitem[{\citenamefont{Mostafazadeh}(2002)}]{Mostafazadeh2002}
\bibinfo{author}{\bibfnamefont{A.}~\bibnamefont{Mostafazadeh}},
  \bibinfo{journal}{J. Math. Phys.} \textbf{\bibinfo{volume}{43}},
  \bibinfo{pages}{205} (\bibinfo{year}{2002}).

\bibitem[{\citenamefont{Mostafazadeh}(2010)}]{Mostafazadeh2010}
\bibinfo{author}{\bibfnamefont{A.}~\bibnamefont{Mostafazadeh}},
  \bibinfo{journal}{Int. J. Geom. Methods Mod. Phys.}
  \textbf{\bibinfo{volume}{07}}, \bibinfo{pages}{1191} (\bibinfo{year}{2010}).

\bibitem[{\citenamefont{Bender}(2007)}]{Bender2007}
\bibinfo{author}{\bibfnamefont{C.~M.} \bibnamefont{Bender}},
  \bibinfo{journal}{Rep. Prog. Phys.} \textbf{\bibinfo{volume}{70}},
  \bibinfo{pages}{947} (\bibinfo{year}{2007}).

\bibitem[{\citenamefont{Das}(2011)}]{Das2011}
\bibinfo{author}{\bibfnamefont{A.}~\bibnamefont{Das}}, \bibinfo{journal}{J.
  Phys.: Conf. Ser.} \textbf{\bibinfo{volume}{287}}, \bibinfo{pages}{012002}
  (\bibinfo{year}{2011}).

\bibitem[{\citenamefont{Zhang et~al.}(2020{\natexlab{a}})\citenamefont{Zhang,
  Qin, and Xiao}}]{Zhang2020}
\bibinfo{author}{\bibfnamefont{R.}~\bibnamefont{Zhang}},
  \bibinfo{author}{\bibfnamefont{H.}~\bibnamefont{Qin}}, \bibnamefont{and}
  \bibinfo{author}{\bibfnamefont{J.}~\bibnamefont{Xiao}}, \bibinfo{journal}{J.
  Math. Phys.} \textbf{\bibinfo{volume}{61}}, \bibinfo{pages}{012101}
  (\bibinfo{year}{2020}{\natexlab{a}}).

\bibitem[{\citenamefont{Guo et~al.}(2020)\citenamefont{Guo, Hou, He, and
  Chien}}]{Guo2020}
\bibinfo{author}{\bibfnamefont{H.}~\bibnamefont{Guo}},
  \bibinfo{author}{\bibfnamefont{X.-Y.} \bibnamefont{Hou}},
  \bibinfo{author}{\bibfnamefont{Y.}~\bibnamefont{He}}, \bibnamefont{and}
  \bibinfo{author}{\bibfnamefont{C.~C.} \bibnamefont{Chien}},
  \bibinfo{journal}{Phys. Rev. B} \textbf{\bibinfo{volume}{101}},
  \bibinfo{pages}{104310} (\bibinfo{year}{2020}).

\bibitem[{\citenamefont{Hou et~al.}(2021)\citenamefont{Hou, Guo, and
  Chien}}]{Hou2021}
\bibinfo{author}{\bibfnamefont{X.-Y.} \bibnamefont{Hou}},
  \bibinfo{author}{\bibfnamefont{H.}~\bibnamefont{Guo}}, \bibnamefont{and}
  \bibinfo{author}{\bibfnamefont{C.~C.} \bibnamefont{Chien}},
  \bibinfo{journal}{Phys. Rev. A} \textbf{\bibinfo{volume}{104}},
  \bibinfo{pages}{023303} (\bibinfo{year}{2021}).

\bibitem[{\citenamefont{Morachis~Galindo
  et~al.}(2021)\citenamefont{Morachis~Galindo, Rojas, and
  Maytorena}}]{Galindo2021}
\bibinfo{author}{\bibfnamefont{D.}~\bibnamefont{Morachis~Galindo}},
  \bibinfo{author}{\bibfnamefont{F.}~\bibnamefont{Rojas}}, \bibnamefont{and}
  \bibinfo{author}{\bibfnamefont{J.~A.} \bibnamefont{Maytorena}},
  \bibinfo{journal}{Phys. Rev. A} \textbf{\bibinfo{volume}{103}},
  \bibinfo{pages}{042221} (\bibinfo{year}{2021}).

\bibitem[{\citenamefont{Zhang et~al.}(2020{\natexlab{b}})\citenamefont{Zhang,
  Qin, and Xiao}}]{Zhang2021}
\bibinfo{author}{\bibfnamefont{R.}~\bibnamefont{Zhang}},
  \bibinfo{author}{\bibfnamefont{H.}~\bibnamefont{Qin}}, \bibnamefont{and}
  \bibinfo{author}{\bibfnamefont{J.}~\bibnamefont{Xiao}}, \bibinfo{journal}{J.
  Math. Phys.} \textbf{\bibinfo{volume}{61}}, \bibinfo{pages}{012101}
  (\bibinfo{year}{2020}{\natexlab{b}}).

\bibitem[{\citenamefont{Hou et~al.}(2026)\citenamefont{Hou, Wang, and
  Guo}}]{Hou2026UhlmannQuasiHermitian}
\bibinfo{author}{\bibfnamefont{X.-Y.} \bibnamefont{Hou}},
  \bibinfo{author}{\bibfnamefont{X.}~\bibnamefont{Wang}}, \bibnamefont{and}
  \bibinfo{author}{\bibfnamefont{H.}~\bibnamefont{Guo}},
  \emph{\bibinfo{title}{Theory of the {U}hlmann phase in {Q}uasi-{H}ermitian
  quantum systems}} (\bibinfo{year}{2026}), \eprint{2603.01908},
  \urlprefix\url{https://arxiv.org/abs/2603.01908}.

\bibitem[{\citenamefont{Uhlmann}(1986)}]{Uhlmann86}
\bibinfo{author}{\bibfnamefont{A.}~\bibnamefont{Uhlmann}},
  \bibinfo{journal}{Rep. Math. Phys.} \textbf{\bibinfo{volume}{24}},
  \bibinfo{pages}{229} (\bibinfo{year}{1986}).

\bibitem[{\citenamefont{Wang et~al.}(2010)\citenamefont{Wang, Chia, and
  Zhang}}]{Wang_2010}
\bibinfo{author}{\bibfnamefont{Q.-h.} \bibnamefont{Wang}},
  \bibinfo{author}{\bibfnamefont{S.-z.} \bibnamefont{Chia}}, \bibnamefont{and}
  \bibinfo{author}{\bibfnamefont{J.-h.} \bibnamefont{Zhang}},
  \bibinfo{journal}{Journal of Physics A: Mathematical and Theoretical}
  \textbf{\bibinfo{volume}{43}}, \bibinfo{pages}{295301}
  (\bibinfo{year}{2010}).

\bibitem[{\citenamefont{Tang et~al.}(2025)\citenamefont{Tang, Hou, and
  Guo}}]{PhysRevB.111.174310}
\bibinfo{author}{\bibfnamefont{J.-C.} \bibnamefont{Tang}},
  \bibinfo{author}{\bibfnamefont{X.-Y.} \bibnamefont{Hou}}, \bibnamefont{and}
  \bibinfo{author}{\bibfnamefont{H.}~\bibnamefont{Guo}},
  \bibinfo{journal}{Phys. Rev. B} \textbf{\bibinfo{volume}{111}},
  \bibinfo{pages}{174310} (\bibinfo{year}{2025}),
  \urlprefix\url{https://link.aps.org/doi/10.1103/PhysRevB.111.174310}.

\bibitem[{\citenamefont{Wang et~al.}(2026)\citenamefont{Wang, Hou, and
  Guo}}]{Wang2026Obstructions}
\bibinfo{author}{\bibfnamefont{M.-Z.} \bibnamefont{Wang}},
  \bibinfo{author}{\bibfnamefont{X.-Y.} \bibnamefont{Hou}}, \bibnamefont{and}
  \bibinfo{author}{\bibfnamefont{H.}~\bibnamefont{Guo}},
  \emph{\bibinfo{title}{Geometric and topological obstructions to
  hermitianization in {Q}uasi-{H}ermitian quantum systems}}
  (\bibinfo{year}{2026}), \eprint{2605.05335},
  \urlprefix\url{https://arxiv.org/abs/2605.05335}.

\end{thebibliography}
\end{document}